\documentclass[traditionalabstract]{aa}

\usepackage{graphicx}
\usepackage{txfonts}
\usepackage[]{hyperref} 
\usepackage{caption}
\usepackage{float}
\usepackage[]{natbib}

\usepackage{xcolor}
\hypersetup
{
    colorlinks,
    linkcolor={red!50!black},
    citecolor={blue!70!black},
    urlcolor={blue!80!black}
}

\makeatletter
\renewcommand*\aa@pageof{, page \thepage{} of \pageref*{LastPage}}
\makeatother

\usepackage{placeins}

\usepackage{subcaption}
\usepackage{graphicx}
\usepackage{multirow}
\usepackage{tikz,pgfplots}
\usepackage{colortbl}

\begin{document}

   \date{Received xxx XXX, xxx; accepted xxx, XXX}

   \title{A universal chromosome map for globular clusters: chemical calibration and environmental regulation of the multiple populations}

\author{C. Lardo\inst{1,2}
\and M. Salaris\inst{2,3}
\and N. Bastian\inst{4}
\and C. Charbonnel\inst{5, 6}
\and E. Dalessandro\inst{2}
\and E. Dondoglio\inst{7}
\and M. Gieles\inst{8, 9, 10}
\and M. G. H. Krause\inst{11}
\and F. Martins\inst{12}
\and S. Monty\inst{13}
  }

\institute{
Dipartimento di Fisica e Astronomia, Universit\`a degli Studi di Bologna, Via Gobetti 93/2, I-40129 Bologna, Italy
\and
INAF - Osservatorio di Astrofisica e Scienza dello Spazio di Bologna, via Piero Gobetti 93/3, I-40129 Bologna, Italy
\and
 Astrophysics Research Institute, Liverpool John Moores University, 146 Brownlow Hill, Liverpool L3 5RF, UK
\and 
Donostia International Physics Center (DIPC), Paseo Manuel de Lardizabal, 4, 20018, Donostia-San Sebastián, Guipuzkoa, Spain; IKERBASQUE, Basque Foundation for Science, 48013, Bilbao, Spain
\and
Department of Astronomy, University of Geneva, CH-1290 Versoix, Switzerland
\and
IRAP, UMR 5277, CNRS and Université de Toulouse, 14 Av. E. Belin, 31400, Toulouse, France
\and
Physics Department, American University of Sharjah, P.O. Box 26666, Sharjah, UAE
\and
ICREA, Pg. Lluís Companys 23, E-08010 Barcelona, Spain
\and 
Institut de Ciències del Cosmos (ICCUB), Universitat de Barcelona (UB), c. Martí i Franquès, 1, E-08028 Barcelona, Spain 
\and 
Institut d'Estudis Espacials de Catalunya (IEEC), Edifici RDIT, Campus UPC, E-08860 Castelldefels (Barcelona), Spain
\and
Centre for Astrophysics Research, Department of Physics, Astronomy and Mathematics, University of Hertfordshire, College Lane, Hatfield, Hertfordshire, UK
\and 
LUPM, Université de Montpellier, CNRS, Place Eugène Bataillon, 34095, Montpellier, France
\and
Institute of Astronomy, University of Cambridge, Madingley Road, Cambridge CB3 0HA, UK; Department of Astronomy, New Mexico State University, Las Cruces, NM 88003, USA
}

   \date{Received September xx, xxx; accepted March xx, xxxx}

\abstract
{}
{Chromosome maps (ChMs) are two-dimensional diagrams of pairs of pseudocolours built from  
UV and optical photometry, widely used as powerful diagnostics of the multiple stellar populations 
phenomenon in massive star clusters. Their raw morphology is, however, affected by
the metallicity-dependent photometric response of the filters employed, which prevents an 
unbiased comparison across cluster samples spanning a range of metallicities.
We aim at identifying a cross-cluster ChM framework that fully accounts for this dependence, 
to allow for an unbiased investigation of the physical drivers of the 
multiple-population diversity.}
{We analyse ChMs for 23 Galactic globulars, and 
devise a technique to correct the individual raw maps for the effect of the clusters' different metallicities. On the final \lq\lq universal\rq\rq~ ChM, we define a new photometric enrichment 
index \(S_{\rm ChM,z}\) that we validate using APOGEE spectroscopy 
of clusters' stars.
We then compare the values of this 
index for our sample of globulars with their masses, structural parameters, orbital quantities, and
accretion-origin classifications.}
{We find that the values of \(S_{\rm ChM,z}\) correlate with the multivariate chemical abundance ranges of the enriched population 
and with the aluminium spread. 
In the full sample, \(S_{\rm ChM,z}\) increases with initial
mass but is most strongly correlated with a covariant family of
orbital-confinement quantities (\(z_{\max}\), vertical action, apocentre,
Galactocentric radius, and orbital energy).}
{The corrected ChMs provide a chemically meaningful, population-level measure of
the enriched sequence. The \(S_{\rm ChM,z}\) index does not trace any single
abundance ratio, but captures the cluster-to-cluster amplitude of the combined
light-element variations, with particular sensitivity to the high-temperature
Mg--Al/O component of proton-capture processing. Its dependence on both cluster
potential depth and orbital confinement suggests that the extent of 2P chemical
diversity is shaped by internal enrichment physics together with an
environmental imprint, whether inherited at formation, modified by early
evolution, or partly filtered by subsequent survival.
}

\keywords{
globular clusters: general ---
Stars: abundances ---
Stars: Population II ---
Galaxy: kinematics and dynamics ---
Techniques: photometric ---
Techniques: spectroscopic
}

\titlerunning{A universal chromosome map for globular clusters}
\authorrunning{Lardo et al.}
\maketitle

\section{Introduction}
\label{sec:intro}

Galactic globular clusters (GCs) host multiple stellar populations
(MPs) characterised by star-to-star abundance variations in light
elements following proton-capture patterns: enhanced N and Na, depleted
C and O, variations in He, and, in some clusters, Mg--Al variations
\citep[e.g.,][]{gratton12,bl18,gratton19}. Heavier elements are
generally homogeneous within individual clusters
\citep{dacosta16,marino19}. The origin of these patterns remains a major
open problem. Proposed enrichment channels include asymptotic giant
branch stars, fast-rotating massive stars, massive interacting binaries,
very massive stars, and supermassive stars, embedded in different
cluster-formation scenarios
\citep[e.g.,][]{Ventura2001,Maeder2006,Prantzos2006,Decressin07,Demink2009,Vink2018,Denissenkov2014,decressin_scenario,dercole08,dantona16,bastian:13,Elmegreen2017,gieles18,gieles25,renzini22}.
Nonetheless, a unifying physical description has not yet emerged, partly because
observational diagnostics have not always provided chemically validated
cross-cluster measures of enrichment strength.

The chromosome map (ChM) introduced by \citet{milone:15,milone17} was a
major step forward. By combining ultraviolet-sensitive pseudo-colours
from high-precision \emph{HST} photometry, it separates primordial (1P)
and enriched (2P) stars in a two-dimensional diagram, and enables a
homogeneous description of MPs across large cluster samples
\citep{piotto15}. Subsequent work has shown that different regions of
the ChM correspond to distinct abundance patterns
\citep[e.g.,][]{marino19,lardo23,legnardi22,carretta24,carretta25,dondoglio25}, and
that its extension correlates with both cluster mass \citep{milone17} and helium variations \citep{milone18}.
The ChM has therefore become one of the primary empirical
diagnostics of the MP phenomenon.

However, the ChM is still a phenomenological construct. Its
zero-point and scale depend on metallicity, because the ultraviolet
filters respond differently to fixed variations in He, C, N, and O at
different \([\mathrm{Fe/H}]\) \citep[e.g.,][]{marino19,carretta24,carretta25}.
The measured 1P--2P separations and enriched-sequence extensions in the ChM mix intrinsic abundance differences with a metallicity-dependent photometric response, preventing direct comparisons of the enrichment extent
across clusters. Several studies have addressed parts of this problem. \citet{milone18}
interpreted the ChM displacements in terms of He, C, N, O, and Mg variations
through a comparison with synthetic spectra, and showed that the photometric
response to a fixed light-element pattern is metallicity-dependent (see their Fig.~8).  
\citet{marino19} introduced metallicity-dependent scale factors and
demonstrated correlations between normalised ChM coordinates and Na, Al,
O, and Mg abundances. At the same time, spectroscopic comparisons have
shown that the mapping between photometric displacement and individual
abundance ratios is not one-to-one: \citet{carretta24} found mismatches
between photometric and spectroscopic population labels, especially at
low metallicity, and \citet{carretta25} showed that the apparent 1P width
is itself affected by metallicity-dependent UV response. Thus, the ChM is
chemically meaningful, but its raw morphology cannot be interpreted as a
universal abundance scale.

The gap addressed here is therefore twofold. First, the raw ChM must be
corrected for the dominant metallicity-dependent photometric response in order to 
preserve the genuine cluster-to-cluster differences in the
2P morphology. Second, the corrected morphology must be
validated as a population-level tracer of the multivariate proton-capture
diversity of the enriched population, rather than interpreted as a direct
proxy for any single abundance ratio. On this basis we define a corrected 2P
ChM-strength index and use it to investigate how the strength of the
MP phenomenon depends on cluster mass, structure, orbital
properties, and accretion history.

The paper is organised as follows. Section~\ref{sec:data} presents the data. Section~\ref{sec:analysis} introduces the metallicity correction,
defines the corrected ChM-extension index, and validates it with APOGEE
abundances. Section~\ref{sec:discussion} discusses the physical drivers of the index, and
Section~\ref{sec:summary} summarises the conclusions.

\section{Data and clusters' sample}
\label{sec:data}

\subsection{Photometric data and chromosome maps}
\label{sec:gcs}

We analyse the HST photometry from the UV Legacy Survey of Galactic Globular
Clusters \citep[HUGS;][]{piotto15,nardiello18}, which provides homogeneous
ultraviolet and optical photometry for 56 Galactic GCs. From this parent
sample we selected 23 clusters suitable for a homogeneous cross-cluster
comparison of the ChM morphology (see below). For these clusters, we compiled a
homogeneous set of cluster parameters, including metallicity, present-day mass
(\(M_{\rm cur}\)), inferred initial mass (\(M_{\rm ini}\)), half-light and
half-mass radii, and escape velocity (\(v_{\rm esc}\)), taken from
\citet{harris} and \citet{baumgardt18}.
Our sample spans the ranges $-2.37 \leq [\mathrm{Fe/H}] \leq -0.44$, $5.46 \leq \log(M_{\rm ini}/M_\odot) \leq 6.41$, and $ 4.78 \leq \log(M_{\rm cur}/M_\odot) \leq 5.93$. The corresponding escape velocities range from $v_{\rm esc} \simeq 10.8~{\rm km~s^{-1}}$ to $v_{\rm esc} \simeq 53.1~{\rm km~s^{-1}}$. 
The clusters also probe a broad range of Galactic
environments, from inner, spatially confined systems to clusters reaching the
intermediate halo, with maximum vertical excursions of
\(z_{\max}\simeq1\)--25 kpc and apocentric radii of
\(r_{\rm apo}\simeq3\)--30 kpc. 

The sample therefore covers a relatively broad range in
metallicity, mass, structural parameters, potential depth, and orbital
confinement, and provides useful leverage for testing whether corrected ChM
morphology depends on Galactic environment in addition to the mass--metallicity
baseline sampled by the HUGS catalogue. We did not, however, attempt to analyse every HUGS
cluster. We excluded systems with ChMs compromised by
strong reddening, by complex Type-II/CNO-enhanced sub-populations  which obscure
the standard 1P--2P morphology \citep[e.g., M~22, NGC~5286][]{dondoglio25}, or by poorly populated red giant branch (RGB) samples preventing a robust measurement of the ChM morphology (e.g., NGC~6838). The final sample is therefore optimised for homogeneous cross-cluster comparison rather
than completeness. 

Stars are selected using the homogeneous cleaning procedure described in
Appendix~\ref{app:phot_cleaning}, which removes field contamination and
mitigates the impact of unresolved binaries, since both effects can broaden or distort the ChM morphology \citep[e.g.][]{marino3201,martins2020}. We restrict the analysis to RGB stars in the luminosity interval where the ChM coordinates are expected to retain the strongest memory of the MP-related light-element pattern. In particular, although selected stars have already undergone first
dredge-up, we avoid including stars above the RGB bump, where additional mixing can
further modify the surface C and N abundances and potentially blur the
photometric separation between primordial and enriched populations.
\citep{gratton00,Placco2014,masseron19,salaris20}. 

The ChM is built from two RGB pseudo-colours, \(\Delta (F275W,F814W)\) and
\(\Delta C_{F275W,F336W,F438W}\), constructed following \citet{milone17}.
For each cluster we trace the red and blue fiducial lines that bracket the RGB
in the relevant colour--magnitude (CMD) and pseudo-CMD planes, and for
every star we measure its colour distance from these fiducials at fixed
\({F814W}\) magnitude. Normalising this distance by a reference RGB width and rotating
the axis so that the fiducials become vertical---the operation commonly
referred to as \lq\lq verticalising\rq\rq---removes the intrinsic slope and curvature
of the RGB. The two resulting coordinates therefore isolate the
star-to-star colour spread driven by chemical differences (He, C, N, O) from
the trivial spread set by evolutionary position along the RGB \citep[see, e.g.,][for a review]{cassisi20}. By construction,
1P stars cluster near \(\Delta_{F275W,F814W}\simeq0\), while N-enriched 2P stars
are displaced to more negative \(\Delta_{F275W,F814W}\) and higher
\(\Delta C_{F275W,F336W,F438W}\).

\subsection{Population assignment}
\label{sec:pop_assignment}

We model each ChM as a Gaussian Mixture Model (GMM), fitting
up to \(K=5\) components and selecting \(K\) by minimising the Bayesian Information
Criterion (BIC), which penalises additional components to avoid over fitting
sparse maps. We use the GMM as a flexible, homogeneous descriptor of the
stellar density distribution. In well-characterised clusters
the individual components are known to correspond, at least
in part, to distinct chemical sub-populations \citep[e.g.][]{carretta15,milone:15}. However, since the optimal
number of components K selected by BIC varies across the
sample, we do not enforce this correspondence at the
component level, and retain only the coarse 1P/2P partition
used throughout this
work: the component whose centroid lies closest to the field-like locus
(\(\Delta_{F275W,F814W}\simeq0\), low \(\Delta C_{F275W,F336W,F438W}\)) is assigned to 1P, and all
remaining components are merged into 2P. We note that star assignment to each Gaussian component is probabilistic
and follows the posterior probability under the fitted
mixture, which incorporates the full covariance of each
component and therefore the intrinsic spread and
observational errors of the underlying distribution. The
mapping from components to 1P/2P labels is instead
deterministic and operates at the component level, as
described above.
This procedure yields a 1P/2P
classification that is independent of the number of components needed to
describe the internal morphology of any given cluster, and is therefore
uniform across the sample regardless of ChM complexity. 
Details of the photometric classification, including the choice of pseudo-colour axes and the treatment of clusters with anomalous sub-populations, are given in
Appendix~\ref{app:phot_assignment}.

The spectroscopic validation in Sect.~\ref{sec:chemical_validation} is
performed independently. There, 1P and 2P stars are classified using APOGEE
abundances \citep{schiavon24} with an extreme-deconvolution Gaussian mixture model \citep[XDGMM;][]{Bovy2011} in the
five-dimensional abundance space
\([\mathrm{C/Fe}], [\mathrm{N/Fe}], [\mathrm{O/Fe}], [\mathrm{Mg/Fe}],
[\mathrm{Al/Fe}]\), including the individual abundance uncertainties. This
separation between photometric morphology and spectroscopic classification
prevents the chemical calibration from being circular. Details are provided in Appendix~\ref{app:chem_classification}.

\begin{figure*}
\centering
\includegraphics[width=\textwidth]{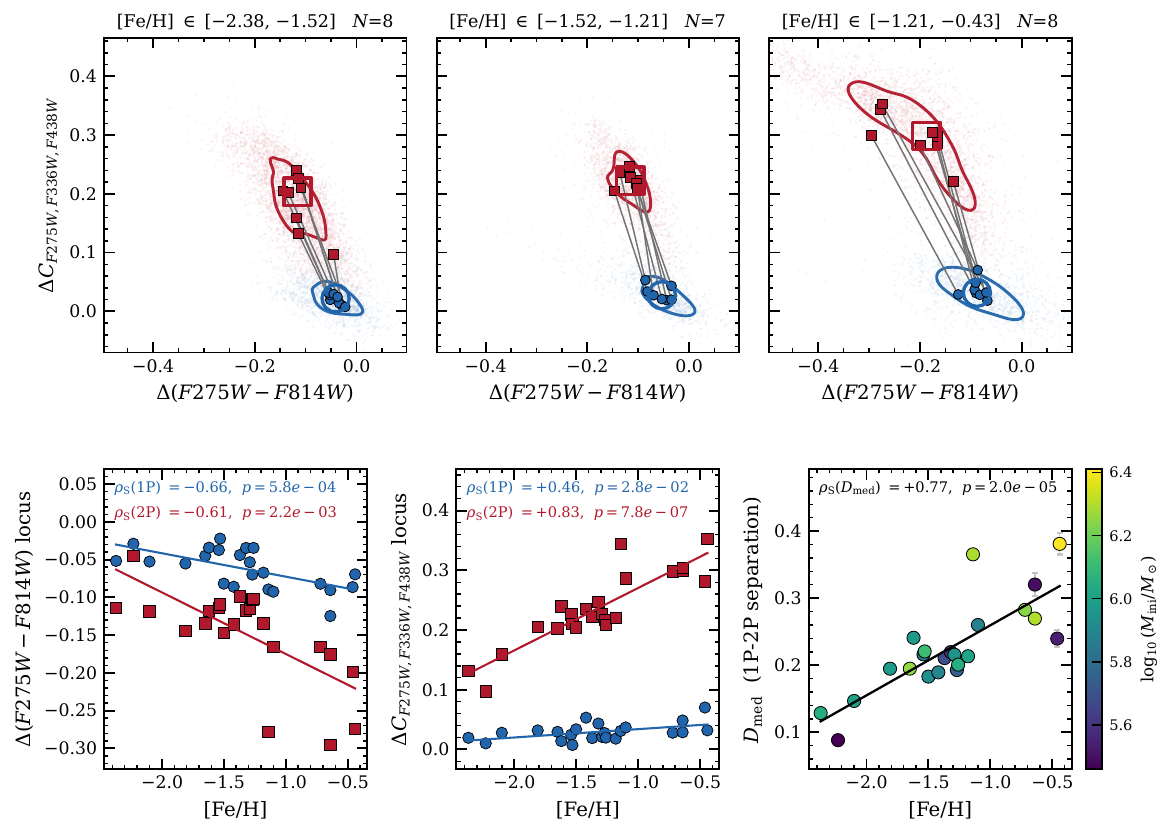}

\caption{Combined raw ChMs and their metallicity dependence.
The top panels show the raw ChMs separated into three bins of increasing
metallicity, from left to right (see top of each panel; \(N\) is the number of
clusters in each group). The metallicity-bin boundaries were chosen
empirically to divide the cluster sample into three approximately equally
populated groups; they are used only for visualisation and do not represent
physically motivated metallicity thresholds.
All panels are displayed on the same coordinate scale. 
Individual stars are shown as faint points for the 1P (blue) and 2P (red)
samples. Coloured contours show the aggregate KDE density of each population
within each metallicity bin, enclosing 39.3 
per cent of the total
probability. This is the 1$\sigma$-equivalent region in two
dimensions. 
Filled symbols mark the 1P and 2P centroids of individual clusters, connected
by grey segments that indicate the raw median 1P--2P separation,
\(D_{\rm med}\). Large open symbols show the median 1P (circles) and 2P
(squares) loci in each metallicity bin.
The bottom panels quantify the differences among these maps cluster by
cluster: the left and middle panels show the metallicity dependence of the 1P
and 2P centroid positions in \(\Delta(F275W-F814W)\) and
\(\Delta C_{F275W,F336W,F438W}\), respectively, while the right panel shows the
raw median 1P--2P separation as a function of metallicity. Points in this panel
are also colour-coded by initial cluster mass. Solid lines show linear fits,
and the quoted Spearman coefficients quantify the monotonic trends.}

\label{fig:raw_metallicity_dependence}
\end{figure*}

\section{Analysis}\label{sec:analysis}

\subsection{The raw chromosome map and its metallicity dependence}
\label{sec:master_map}

Before using the ChM morphology as a cluster-to-cluster diagnostic, it is
necessary to assess whether the raw ChM coordinates can be compared directly
across the full metallicity range of the photometric sample. The ChM is built from ultraviolet
and blue HST colours that are sensitive to He, C, N, and O abundance variations \citep{piotto15},
but the photometric response to a fixed abundance pattern is itself a function
of the stellar atmospheric parameters \citep{sbordone11,salaris20}. In particular, at higher metallicity, the
molecular bands in the wavelength range of the $F275W$, $F336W$, and $F438W$ filters affect the pseudo-colours more substantially. Consequently, the same intrinsic
light-element variation produces a different displacement in the raw
\(\Delta(F275W-F814W)\) versus
\(\Delta C_{F275W,F336W,F438W}\) plane depending on
\([\mathrm{Fe/H}]\) \citep[e.g.][]{milone18, lardo18, marino19,carretta25}. 

Figure~\ref{fig:raw_metallicity_dependence} highlights this effect.
The top panels show the raw ChMs of all the 23 clusters in the main raw-ChM sample, divided into three bins of increasing metallicity and displayed
on the same coordinate scale. In each bin, the 1P stars form a relatively compact
locus, while the enriched population is displaced toward larger
\(\Delta C_{F275W,F336W,F438W}\) and more negative
\(\Delta(F275W-F814W)\), from the metal-poor to the metal-rich regime.
Thus, even before any quantitative modelling, the raw ChMs
reveal a systematic metallicity dependence: although the 1P
reference locus is shared across the sample by construction
\citep{milone:17}, the observed 1P and 2P centroids drift
with [Fe/H] within this common frame, so raw ChM coordinates
cannot be compared directly across clusters as amplitude
measurements.

The lower panels of Fig.~\ref{fig:raw_metallicity_dependence} quantify this
behaviour cluster by cluster. In the raw ChM plane,
\(x\equiv\Delta(F275W-F814W)\) and
\(y\equiv\Delta C_{F275W,F336W,F438W}\). For each cluster, we measured the median 1P locus
\((x_{\rm 1P},y_{\rm 1P})\), the median 2P locus
\((x_{\rm 2P},y_{\rm 2P})\), and the raw median separation
\[
    D_{\rm med}
    =
    \left[
    (x_{\rm 2P}-x_{\rm 1P})^2
    +
    (y_{\rm 2P}-y_{\rm 1P})^2
    \right]^{1/2}.
\]
The lower left-hand and middle panels show that both the 1P and 2P centroid positions vary
systematically with \([\mathrm{Fe/H}]\), in both raw ChM coordinates. The right-hand panel
shows that the apparent 1P--2P separation also changes with metallicity. This is
important because it implies that the metallicity dependence is not simply a
global translation of the entire map: it also affects the apparent amplitude of
the enriched sequence, although this separation may also include intrinsic
cluster-to-cluster variations in the enriched-population extent. Therefore, a raw separation such as \(D_{\rm med}\) cannot
be interpreted directly as an intrinsic enrichment strength.

\subsection{Empirical model for the raw ChM loci}

To describe the metallicity-dependent drift of the raw ChM, we model the
cluster-level ChM loci as empirical functions of metallicity and two nuisance
cluster parameters. Again, we denote the raw ChM coordinates as
\(x\equiv\Delta(F275W-F814W)\) and \(y\equiv\Delta C_{F275W,F336W,F438W}\). 
For each cluster, we measure the
median 1P and 2P loci, \((x_{\rm 1P},y_{\rm 1P})\) and \((x_{\rm 2P},y_{\rm 2P})\), and a 2P-tail
anchor, \((x_{\rm 2P,max},y_{\rm 2P,max})\). The latter is not the position of
the single most extreme star, but the 95th percentile of the 2P distribution
projected along the direction from the 1P median locus to the 2P median locus.
It is therefore a robust measure of the high-enrichment end of the raw 2P
sequence. For each coordinate
\[
    q \in
    \{x_{\rm 1P},y_{\rm 1P},x_{\rm 2P},y_{\rm 2P},x_{\rm 2P,max},y_{\rm 2P,max}\},
\]
we fit
\[
    q_i = \alpha + \beta_{\rm [Fe/H]}[\mathrm{Fe/H}]_i
         + \beta_M \log M_{\rm ini} + \beta_E E(B-V)_i + \epsilon_i ,
\]
with an intrinsic-scatter term \(\epsilon_i\)\footnote{A linear term in
\([\mathrm{Fe/H}]\) is sufficient over the sampled range: the quadratic
coefficient is consistent with zero, and a linear model is preferred by
leave-one-out cross-validation.}. The metallicity term captures the
photometric response we want to remove; the mass term accounts for the known
mass dependence of the MP extent; and the reddening term absorbs residual colour
systematics. The goal is not to infer an abundance scale from the raw
coordinates, but to identify the first-order trends that would otherwise bias a
cross-cluster comparison.

The dominant term is metallicity, with the clearest effect on the 2P
pseudo-colour ordinate \(y=\Delta C_{F275W,F336W,F438W}\), which responds to the
CN and NH molecular bands that carry the MP signal. For the median 2P locus and
the most extended 2P point we find
\[
    \beta_{[\mathrm{Fe/H}]}(y_{\rm 2P}) = +0.10 \pm 0.01~{\rm mag~dex^{-1}},
\]
\[
    \beta_{[\mathrm{Fe/H}]}(y_{\rm 2P,max}) = +0.12 \pm 0.02~{\rm mag~dex^{-1}}.
\]
At higher metallicity, the same light-element abundance difference therefore
produces a larger raw pseudo-colour displacement. The 1P loci show weaker trends and lower explained variance
than the 2P loci: this is expected, since the 1P abundance pattern is
comparatively homogeneous at the precision relevant here\footnote{We note in passing that the 1P locus in the corrected frame shows a small but non-zero residual width across the sample, which is not correlated with [Fe/H] or initial cluster mass. This indicates that the 1P sequence is not fully compatible with a chemically homogeneous population, and points to small intrinsic chemical inhomogeneities within the primordial component \citep[e.g.][]{lardo18,marino3201,lardo23,carretta24,carretta25}. A dedicated analysis of the 1P homogeneity is beyond the scope of the present paper: the 1P locus is used throughout only as the local reference against which the 2P morphology is measured.}.

The key physical point is that this drift reflects how the photometry
responds to a fixed abundance pattern, not a change in the abundance pattern
itself. Two independent lines of evidence support this. First, the synthetic
1P--2P pseudo-colour separations of \citet{branco24} increase with metallicity
at \(\simeq+0.125\) mag dex\(^{-1}\), matching our empirical separation slope
(\(+0.098\); 95\% highest-density interval \([0.067,0.130]\) i.e. the Bayesian 95\% credible range) in both sign and amplitude. The same intrinsic chemistry is thus mapped onto a larger
colour separation simply because the photometric response is stronger at higher
metallicity. Second, the APOGEE-based 2P abundance widths confirm from the spectroscopic side that the intrinsic enrichment metrics we introduce in Sect.~\ref{sec:chemical_validation} shows no significant dependence on \([\mathrm{Fe/H}]\),  
and the individual 2P widths in N, O, Mg, and Al show no positive trend. Thus, the APOGEE data do not support a
metallicity trend in the intrinsic 2P chemical width large enough to produce
the raw ChM drift.

We therefore treat the raw \([\mathrm{Fe/H}]\) drift as an essentially photometric effect, to be removed, while preserving the residual 2P extent and width as candidate physical signals.
The mass term falls into this second category. A secondary, weaker dependence on cluster mass
is present mainly for the most extended part of the 2P sequence: at fixed metallicity, more massive clusters
reach more extreme raw ChM positions, consistent with the established mass
dependence of the MP phenomenon \citep{carretta10,milone17}. We deliberately do not remove this term, 
because any residual mass or potential-well
dependence is a candidate physical signal rather than a photometric artefact
(Sect.~\ref{sec:chm_strength}). The reddening term, by contrast, is treated as an observational nuisance.
Its coefficient is relevant mainly along the \(\Delta(F275W-F814W)\) axis, where
including \(E(B-V)\) reduces the intrinsic scatter, as expected for a residual
colour effect. To test whether this term instead reflects a genuine dependence
on Galactic position, we repeated the fits replacing \(E(B-V)\) with the
present-day vertical height \(|Z|\). This substitution does not reduce the
scatter, and in joint fits the \(|Z|\) coefficient is consistent with zero once
\(E(B-V)\) is included. We therefore interpret the colour-axis dependence as
observational in origin, most likely due to residual differential reddening or
small intrinsic variations among 1P stars \citep{marino19,lardo23}, rather
than as a physical correlation with Galactic position.

\subsection{A physically corrected universal ChM frame}
\label{sec:chm_strength}

As discussed above, the apparent ChM amplitude mixes intrinsic chemical
differences with metallicity-dependent photometric amplification. To place all clusters on a common photometric scale, we subtract the
empirical dependence of the 1P and 2P loci on \([\mathrm{Fe/H}]\) and
\(E(B-V)\), using the population- and coordinate-dependent slopes derived
in Sect.~\ref{sec:master_map}. The correction is evaluated relative
to the sample median values \([\mathrm{Fe/H}]_{\rm ref}=-1.32\) and
\(E(B-V)_{\rm ref}=0.05\), and shifts the raw coordinates to the values
they would have at this common metallicity and reddening. 
After correction, the maps
remain in magnitude units and preserve the cluster-to-cluster ChM morphology
that we aim to test. The correction is applied to 12,475 RGB stars in the 23
clusters of the science sample.

\begin{figure}
\centering
\includegraphics[width=0.98\columnwidth]{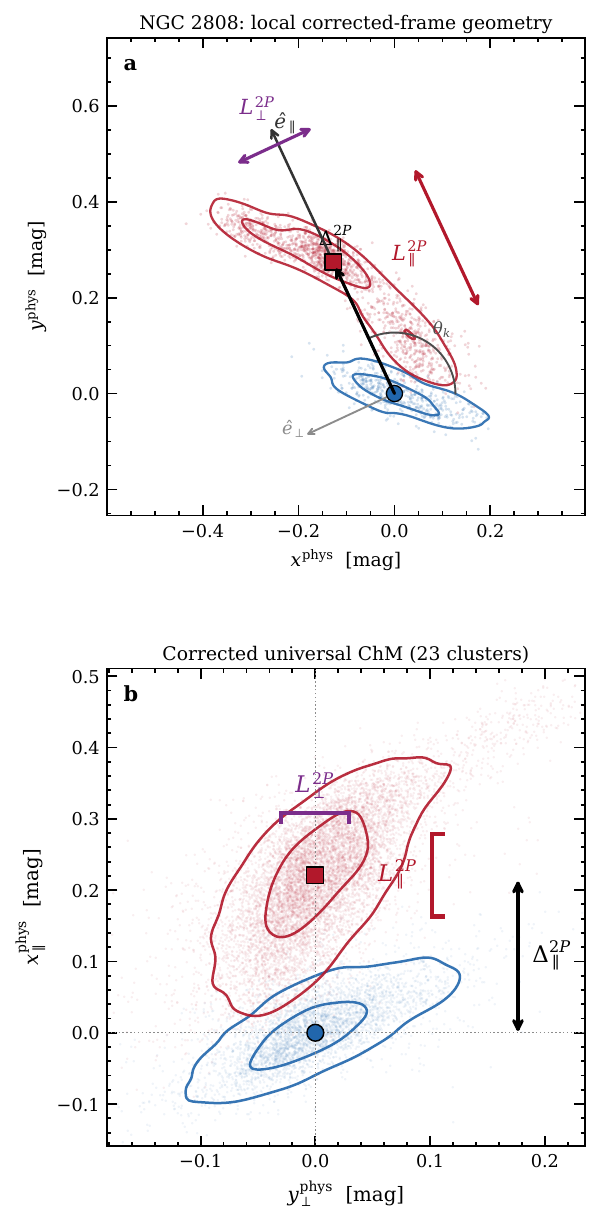}
\caption{Definition of the corrected ChM quantities used in the analysis. Panel (a) shows NGC~2808 after removing the empirical \([\mathrm{Fe/H}]\)- and \(E(B-V)\)-dependent photometric terms and centring the map on the median 1P position. The enrichment axis, \(\hat{e}_{\parallel}\), is defined by the 1P--2P vector; \(\hat{e}_{\perp}\) is the perpendicular direction and \(\theta_k\) its orientation. We define \(\Delta_{\parallel}^{2P}\) as the median 2P displacement along \(\hat{e}_{\parallel}\), while \(L{\parallel}^{2P}\) and \(L_{\perp}^{2P}\) are the \(P_{84}-P_{16}\) widths of the 2P projections along and across this axis. Panel (b) shows all clusters stacked in the corrected universal frame, after rotating them to align the cross-cluster median 1P--2P direction vertically. All quantities are in magnitude units.
}
\label{fig:universal_chm_geometry}
\end{figure}

Figure~\ref{fig:universal_chm_geometry} illustrates how the corrected 2P
observables are defined. For each cluster, we compute the corrected median
positions of the 1P and 2P populations. The vector connecting these centroids
defines the local enrichment axis, \(\hat{e}_\parallel\), and the orthogonal
direction defines \(\hat{e}_\perp\). The 2P stars are then projected onto these
two axes after centring the map on the 1P centroid. All morphology quantities
are therefore one-dimensional offsets or widths measured parallel or
perpendicular to the enrichment direction, rather than radial distances in the
full two-dimensional ChM plane.

Measurements are made cluster by cluster in each cluster's own local frame,
because the orientation of the enrichment axis, denoted by \(\theta_k\) in
Fig.~\ref{fig:universal_chm_geometry}, varies slightly among clusters. This
tilt reflects the fact that the relative contribution of different
light-element variations to the 1P--2P displacement need not be identical at
all metallicities. Using a single universal axis would therefore mix
measurements for clusters whose local enrichment axis is tilted with respect to the sample median.

The analysis deliberately focuses on the enriched 2P sequence. The 1P
population defines only the local origin and the enrichment-axis direction; its
internal width is not included in the primary ChM-strength index. This choice is
motivated by previous work \citep{marino19,lardo23}, by our photometric
cleaning diagnostics (Appendix~\ref{app:phot_cleaning}), and by the measurable
raw metallicity trends of the 1P widths, all of which indicate that the 1P
sequence can retain contributions from small pristine abundance variations,
residual reddening, or other photometric systematics. We therefore treat the 1P
as the local baseline and restrict the science measurements to the 2P
projections.

We define three 2P observables in the corrected ChM, illustrated in panel~(a) of
Fig.~\ref{fig:universal_chm_geometry} for NGC~2808:
\begin{itemize}
    \item \(\Delta_\parallel^{2P}\): the median projected distance of 2P stars
    from the 1P centroid along \(\hat{e}_\parallel\). This measures the mean
    offset of the enriched population from the primordial locus.

    \item \(L_\parallel^{2P}\): the inter-percentile width (\(P_{84}-P_{16}\)) of the
    2P sequence along the enrichment axis. This measures how far the enriched
    sequence extends along the dominant 1P--2P displacement. Since this
    direction is primarily driven by the CN/NH-sensitive pseudo-colour
    response, we interpret it as the main projected enrichment axis, dominated
    by N-rich/C,O-poor chemistry \citep[e.g.][]{lardo18,milone18}. A larger
    value indicates a broader range of enrichment or dilution along this main
    projected direction.

    \item \(L_\perp^{2P}\): the inter-percentile width  (\(P_{84}-P_{16}\))
    perpendicular to the enrichment axis. This quantity is not assigned to a
    single abundance, but captures components that move stars away from the
    dominant N-sensitive direction (i.e. He-sensitive colour shifts and
    high-temperature Mg--Al/O variations; 
    \citealp[e.g.][]{milone18,lardo18,marino19,dondoglio25}).
\end{itemize}

As a validation of the corrected frame, we test these observables for residual
trends with \([\mathrm{Fe/H}]\). 
None of the three primary 2P observables retains a significant metallicity dependence,
confirming that the first-order photometric effect has been removed (Sect.~\ref{sec:master_map}). Panel~(b) of Fig.~\ref{fig:universal_chm_geometry} shows all 23 clusters in a
common diagram, obtained by rotating each corrected map so that the
cross-cluster median enrichment direction is vertical. This panel illustrates
the common corrected morphology of the sample, but all measurements are made in
the local cluster frames defined above.

\subsubsection{A single empirical index to describe the 2P enrichment}
\label{sec:S_index}

The three primary ChM observables of Sect.~\ref{sec:chm_strength} are not independent: they describe
different projected properties of the same enriched sequence, namely, its extent
along the enrichment direction, its perpendicular width, and its median
displacement from the 1P locus. We therefore combine them into a single empirical index through a
principal-component analysis (PCA), defining \(S_{\rm ChM}\) as the first
principal component of the standardized three-dimensional space
\((L_{\parallel}^{2P}, L_{\perp}^{2P}, \Delta_{\parallel}^{2P})\):
\[
    S_{\rm ChM}
    \equiv
    \mathrm{PC1}
    \left(
    L_{\parallel}^{2P},
    L_{\perp}^{2P},
    \Delta_{\parallel}^{2P}
    \right).
\]
The sign of PC1 is chosen so that larger \(S_{\rm ChM}\) values correspond to
more extended enriched sequences; \(S_{\rm ChM,z}\) denotes its standardized
form. The PCA is performed on the 23 clusters with measurements of
\(L_{\parallel}^{2P}\), \(L_{\perp}^{2P}\), and
\(\Delta_{\parallel}^{2P}\), using the corrected projected quantities defined
in Fig.~\ref{fig:universal_chm_geometry}. As expected for positively correlated
morphology measures, the first principal component captures most 
of the variance of the shared cluster-to-cluster variation. PC1 explains \(72.6\%\) of the total variance, while PC2 
accounts for \(21.6\%\). 
The PC1 loadings are all
positive and comparable, with values equal to \(+0.642\), \(+0.541\), and \(+0.544\)
for \(L_{\parallel}^{2P}\), \(L_{\perp}^{2P}\), and
\(\Delta_{\parallel}^{2P}\), respectively. Thus, \(S_{\rm ChM}\) increases for
clusters whose 2P sequence is simultaneously longer along the enrichment
direction, broader perpendicular to it, and more strongly displaced from the 1P
locus.
We therefore interpret \(S_{\rm ChM,z}\) as a compact summary of the overall
2P extension in the corrected ChM, not as evidence that a single physical
parameter describes the MP morphology. 

The index is built from the enriched sequence by construction, but a residual
concern is that it could instead be tracking broadening of the reference 1P
locus --- for example from residual reddening or photometric systematics that
survive the correction. If that were the case, \(S_{\rm ChM,z}\) would
correlate with the 1P parallel width. It does not
(\(\rho_{\rm S}=+0.25\), \(p=0.246\)), confirming that the index is driven by
the morphology of the enriched population rather than by the reference P1
sequence.

\begin{figure*}
\sidecaption
\includegraphics[width=12cm]{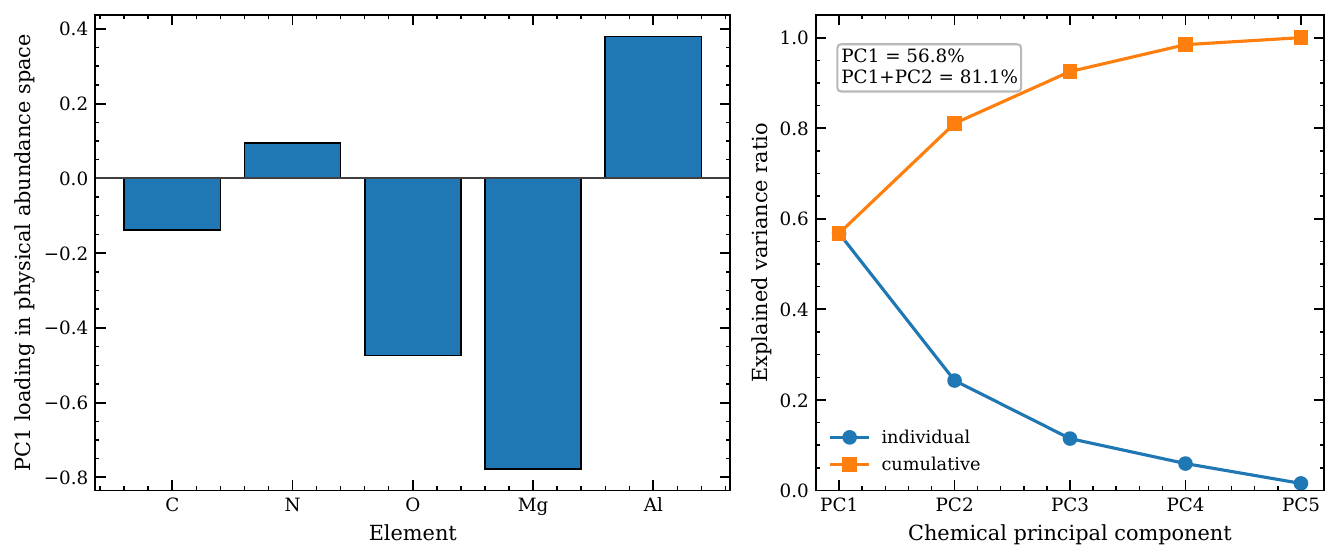}
\caption{Global APOGEE chemical PCA of the reliable 2P sample (see text for details). \textit{Left panel:} PC1 loadings in physical abundance space. The sign convention
is chosen such that positive PC1 corresponds to the proton-capture direction:
enhanced N and Al and depleted C, O, and Mg. The largest loadings are on Mg
and O depletion and Al enhancement, showing that
\(\mathrm{PC1}_{\rm chem}\) is a multivariate high-temperature
proton-capture axis rather than a pure nitrogen coordinate.
\textit{Right panel:} Explained variance ratio of the PCA performed on 648
2P stars in the five-dimensional abundance space.}
\label{fig:apogee_global_pca}
\end{figure*}

\subsubsection{Chemical validation with homogeneous APOGEE abundances}
\label{sec:chemical_validation}

The physically corrected ChM observables of Sect.~\ref{sec:chm_strength} are
photometric measures of the morphology of the 2P. Here we test
their chemical meaning at the cluster level using homogeneous APOGEE abundances
from the value-added catalogue of \citet{schiavon24}. The goal is not to
calibrate ChM coordinates into abundance units, nor to assign a unique abundance
ratio to each photometric displacement --- such a calibration would require a
large sample of stars with both precise ChM positions and homogeneous
spectroscopy of all relevant light elements \citep{marino19,dondoglio25}, which
is not available for this sample. Instead, we ask whether the aggregate
morphology of the corrected 2P sequence traces the internal multivariate
chemical diversity of the enriched population.

We use APOGEE abundances in the five-dimensional light-element space
\(([\mathrm{C/Fe}],[\mathrm{N/Fe}],[\mathrm{O/Fe}],[\mathrm{Mg/Fe}],
[\mathrm{Al/Fe}])\), after quality cuts on membership probability,
signal-to-noise, surface gravity, and abundance flags. Stars are classified into
1P and 2P with an extreme-deconvolution mixture model, and clusters are retained
only when both populations are well sampled. This leaves 11 clusters
with reliable spectroscopic 1P/2P classification overlapping the ChM sample (see  
Appendix~\ref{app:chem_classification}).

The first chemical component, \(\mathrm{PC1}_{\rm chem}\), explains \(56.8\%\)
of the 2P variance (Fig.~\ref{fig:apogee_global_pca}) and shows the canonical
proton-capture pattern: oriented so that positive values denote processing, it
loads positively on Al and N and negatively on Mg, O, and C. The axis is dominated by the
high-temperature Mg--Al--O group, with comparatively weak C and N loadings. The
small N loading is expected: APOGEE samples RGB stars whose surface C and N are
partly affected by evolutionary mixing, diluting the present-day contrast relative
to the birth pattern that drives the UV pseudo-colours
\citep[e.g.][]{salaris20}. \(\mathrm{PC1}_{\rm chem}\) is therefore a data-driven
proton-capture axis, most sensitive here to the Mg--Al/O component.

Having defined a spectroscopic proton-capture axis, we can now test whether the
corrected ChM morphology carries the same chemical information at the cluster
level. We do this in two complementary ways. First, we ask whether the
photometric ChM observables and the APOGEE abundance widths vary together as
part of a common cluster-to-cluster amplitude. To test this, we perform a PCA
on five cluster-level quantities:
\[
\left(
L_\parallel^{2P},
L_\perp^{2P},
\Delta_\parallel^{2P},
{\rm IQR}(\mathrm{PC1}_{\rm chem})_{\rm 2P},
{\rm IQR}([\mathrm{Al/Fe}])_{\rm 2P}
\right).
\]
The first component explains \(85.8\%\) of the total variance, with
positive and comparable loadings on all five quantities. Thus, clusters with
broader corrected ChM sequences also tend to have broader APOGEE
proton-capture distributions, indicating that the photometric and
spectroscopic widths trace the same underlying enriched-population amplitude.
Second, we test the purely photometric index directly against spectroscopic
quantities that were not used to construct it. In the APOGEE overlap
sample, \(S_{\rm ChM,z}\) correlates with both
\({\rm IQR}(\mathrm{PC1}_{\rm chem})_{\rm 2P}\) and
\({\rm IQR}([\mathrm{Al/Fe}])_{\rm 2P}\), with
\(\rho_{\rm S}=+0.61\) and \(p=0.047\) in both cases. Given the small overlap
sample, these correlations should be interpreted as suggestive rather than
definitive. They nevertheless provide a genuine cross-validation: the
photometric index is built only from ChM morphology, yet it tracks independent
spectroscopic measures of the 2P proton-capture spread.

Together, these tests confirm that the corrected ChM morphology is chemically
meaningful at the population level. Clusters with broader corrected enriched
sequences also show broader APOGEE 2P proton-capture distributions. We
therefore interpret \(S_{\rm ChM,z}\) as a photometric tracer of the internal
chemical diversity of the enriched population: not a one-to-one proxy for any
single abundance ratio, but a compact projection of the combined light-element
variations that define the MP phenomenon.

\begin{figure*}
\centering
\includegraphics[width=\textwidth]{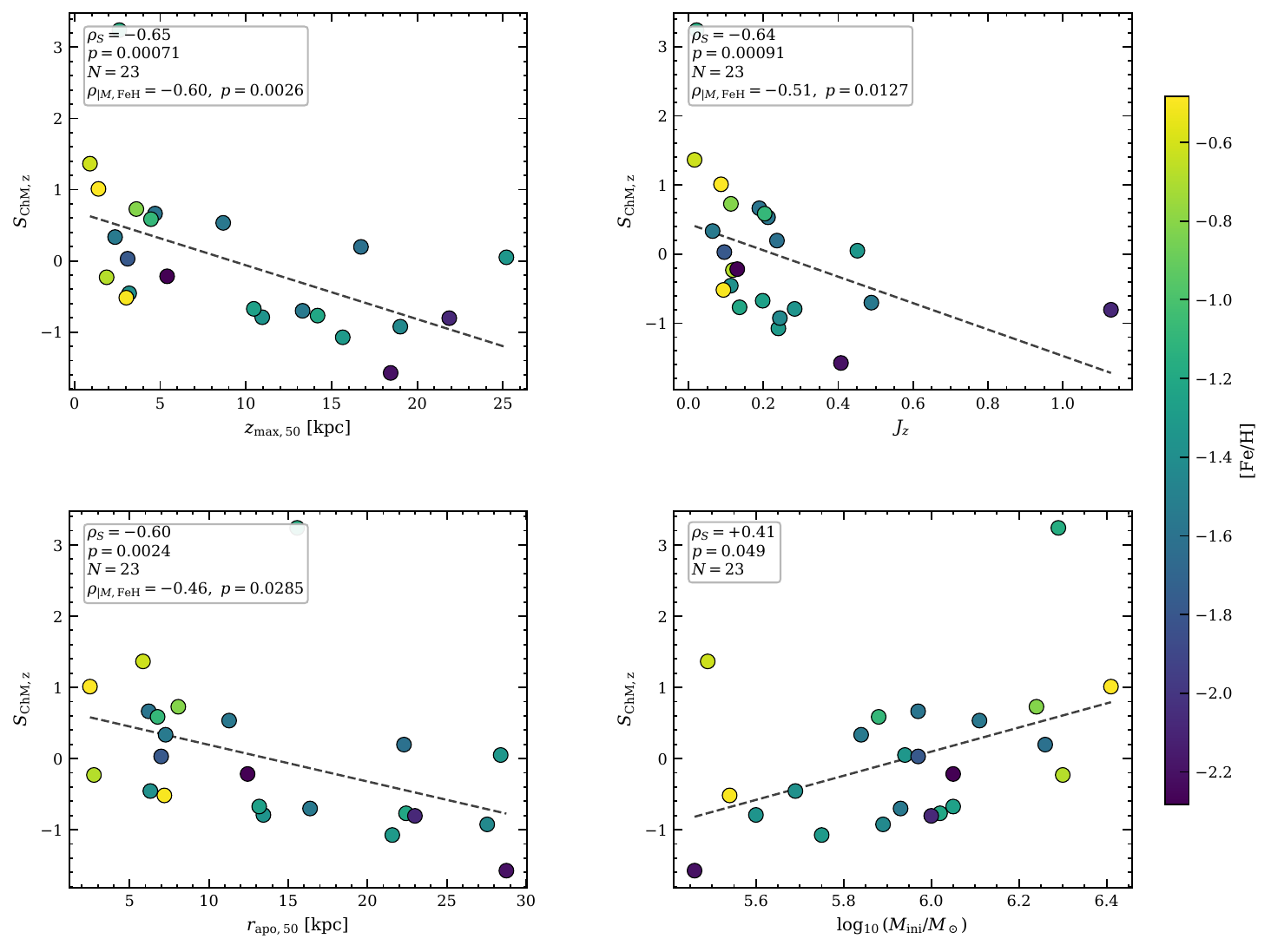}
\caption{Physical drivers of the full-sample ChM-extension index. \textit{From top to bottom, left to right:}\(S_{\rm ChM,z}\) versus maximum orbital height \(z_{\max}\); \(S_{\rm ChM,z}\) versus vertical action \(J_z\); \(S_{\rm ChM,z}\) versus the apocentric distance \(r_{\rm apo, 50}\);  \(S_{\rm ChM,z}\) versus the initial cluster mass. Points are colour coded in terms of 
\([\mathrm{Fe/H}]\), and dashed lines show linear fits for visual guidance. Panels are annotated with the relevant statistics.}
\label{fig:drivers_summary}
\end{figure*}

\section{Discussion}
\label{sec:discussion}

\subsection{Empirical drivers: orbits, mass, and origin labels}
\label{sec:drivers}

We now apply \(S_{\rm ChM,z}\) to the full set of 23 clusters with corrected ChM
measurements and ask which cluster-scale quantities are most closely associated
with the extent of the MP phenomenon. 
For correlations we use the Spearman rank coefficient \(\rho_{\rm S}\)
and assess significance with permutation \(p\)-values
(\(p_{\rm perm}\)), obtained by recomputing \(\rho_{\rm S}\) under \(N_{\rm perm}\)
random reshuffling of one variable\footnote{The subscript ``perm'' is used explicitly in this section to distinguish these probabilities
from the \(p\)-values of other tests used.}. For group comparisons we use the
Mann--Whitney--Wilcoxon rank-sum test \citep{mannwhitney47}, denoted
\(p_{\rm MW}\). Because we test many candidate quantities, we control the
false-discovery rate with the Benjamini--Hochberg procedure
\citep{benjamini95}, reporting adjusted \(q\)-values (\(q_{\rm BH}\)).

\paragraph{Orbit and environment.}
The clearest empirical result of the driver analysis is that the
corrected ChM extension is most strongly linked to orbital confinement. 
The strongest correlations are associated with a set of mutually covariant orbital quantities. We therefore interpret them as different projections of a single orbital-confinement dimension, rather than as independent causal drivers.

Taking the maximum vertical excursion \(z_{\max}\) as a
representative projection of this orbital-confinement dimension, clusters reaching greater heights
above the Galactic plane have less extended corrected ChM morphologies with  \(\rho_{\rm S}(S_{\rm ChM,z},\,z_{\max}) = -0.65,~p=0.001\), as shown in Fig.~\ref{fig:drivers_summary}. Thus, clusters on more
confined orbits host systematically broader enriched-population morphologies.
The same sign and comparable strength hold for the vertical action \(J_z\)
(\(\rho_{\rm S}=-0.64\)), orbital energy (\(-0.61\)), apocentre (\(-0.60\)),
present Galactocentric radius (\(-0.59\)), and the radial action
\(J_r\) (\(-0.52\)), all with \(p_{\rm perm}\) of a few \(\times10^{-3}\). Of all
the cluster parameters tested, only these orbital quantities survive the
Benjamini--Hochberg correction (\(q_{\rm BH}\le0.06\)); no mass, density,
structural, or rotational quantity does\footnote{This statement refers to integrated cluster parameters; more
differential diagnostics of the internal morphology of the enriched population
may carry information that global quantities average out.}.
The
\(z_{\max}\) relation is stable under leave-one-out resampling 
and survives partialling out initial mass and
metallicity jointly (\(\rho_{\rm S}=-0.60\), \(p=0.003\)).  Split-sample tests support the same conclusion. Dividing the clusters at \(z_{\max}=5\) kpc, the orbitally confined group has a median \(S_{\rm ChM,z}\) higher by \(1.3\) standardised units (\(p_{\rm MW}=0.002\)). Consistent differences are found when using apocentre (\(p_{\rm MW}=0.007\)) or Galactocentric radius (\(p_{\rm MW}=0.005\)) as the splitting variable. 
These tests show that the result is not tied to the particular choice
of \(z_{\max}\), but traces a broader orbital-confinement dimension.

These full-sample correlations are, however, potentially entangled with
foreground reddening, which must be addressed before they can be interpreted
physically. Appendix~\ref{app:reddening_robustness} shows that the orbital
signal persists once this is taken into account. Appendix~\ref{app:age}
examines whether the signal could instead reflect baseline chemical differences
or formation epoch, and finds that neither alternative accounts for the
observed trend.

We therefore identify orbital confinement as the dominant external
correlate of the corrected ChM extension: at fixed mass and metallicity,
clusters on inner, spatially confined orbits have more extended enriched
sequences than clusters on higher-energy orbits.

\paragraph{Mass and related quantities.}
Cluster mass plays a real but, in this sample, weaker role. The ChM extension
correlates marginally with initial mass  
($\rho_{\rm S}=+0.41$,
$p=0.049$); see Fig.~\ref{fig:drivers_summary}) but not with present-day mass (\(\rho_{\rm S}=+0.19\),
\(p=0.37\)), consistent with the abundance pattern having been set early, near
the cluster's initial potential well. The weakness of this trend should not be
read as evidence against the well-established mass dependence of the MP
phenomenon \citep[e.g.][]{milone17,milone18,carretta10,schiavon24}. Our sample
spans only \(\sim\!1\) dex in initial mass 
and is concentrated at the high-mass end, where 17 of 23 clusters exceed
\(\log M_{\rm ini}=5.8\). It therefore does not probe the low-mass regime
(\(\log M\lesssim5\)) in which the mass--MP relation is steepest. The clusters
were selected for well-populated, morphologically clean ChMs (Sect.~\ref{sec:gcs}), which
preferentially retains massive systems, so the restricted and top-heavy mass
coverage attenuates any monotonic mass correlation by construction. The orbital
parameters, by contrast, are well sampled across their full range
(\(z_{\max}\simeq1\)--25 kpc), so the comparison between the two is not an
artefact of differential range restriction.  Related quantities (i.e.; velocity
dispersion, density, escape velocity, and core radius) show correlations of
comparable or lower strength that weaken once \(\log M_{\rm ini}\) is controlled,
confirming that they trace the same potential-well scale rather than acting
independently.

The positive sign of the mass trend is qualitatively consistent with several
classes of MP formation scenarios, all of which contain a natural dependence on
cluster mass, although through different physical channels. In some models the
relevant scaling enters through the number of massive polluters or the early
mass budget \citep[e.g.][]{Decressin07,prantzos06}; in others through the
maximum stellar mass, the amount of processed material produced by extremely
massive stars, or collisional self-enrichment during compact cluster formation
\citep[e.g.][]{gieles18,gieles25}; and in retention-based scenarios through the
ability of a deeper potential well to retain and dilute processed ejecta
\citep[e.g.][]{bekki11,bobrick25}. Thus, while the observed mass dependence
does not discriminate among models, it is consistent with the common
expectation that more massive proto-clusters can sustain a larger or more
extended enriched component. This interpretation also connects qualitatively with recent JWST results on
nitrogen-enhanced compact sources at high redshift, which have been discussed
as possible signatures of dense proto-GC-like star formation and
supermassive-star or extremely-massive-star nucleosynthesis
\citep[e.g.][]{charbonnel23,senchyna24,gieles25}. The analogy remains
suggestive, but it points to a common physical regime:
compact, dense star formation in which the production, retention, and dilution
of hot-H-burning material may be especially efficient.

\begin{figure}
\centering
\includegraphics[width=\columnwidth]{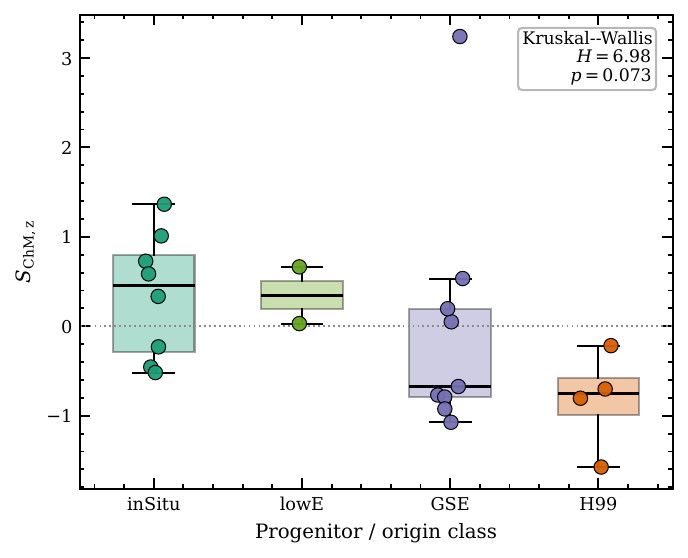}
\caption{Distribution of the full-sample ChM-extension index \(S_{\rm ChM,z}\) by
predicted progenitor/origin class. Points show individual clusters and boxes
show the median and interquartile range for each class.  Labels denote clusters
formed in situ, Gaia--Sausage--Enceladus (GSE), the low-energy Galactic
component (low-E), and the Helmi-stream progenitor (H99). 
}
\label{fig:progenitor_chm_box}
\end{figure}

\paragraph{Candidate progenitor.}
Discrete accretion-class labels are less informative than the continuous orbital
structure (Fig.~\ref{fig:progenitor_chm_box}). Using the probabilistic dynamical
classification of \citet{lardo26}, the median \(S_{\rm ChM,z}\) is highest for
in-situ clusters (\(\tilde{S}_{\rm ChM,z}=+0.46\)), comparable for the low-energy
group (\(+0.35\);~\citealp{massari19,forbes20,kruijssen20}), and markedly lower for Gaia--Sausage--Enceladus
\citep[GSE;][]{helmi18} and Helmi-stream \citep[H99;][]{helmi99} clusters
(\(-0.67\) and \(-0.75\), respectively; tildes denote medians). These differences should be read with caution. The sample contains only two to nine clusters per class, limiting the power of
non-parametric comparisons across the four groups. Accordingly, a
Kruskal--Wallis test \citep{kruskal52} is only marginal
\((p=0.073)\). Collapsing the classification to a binary in-situ/accreted split
yields a significant separation \((p=0.017)\), but the progenitor labels still
do not organise \(S_{\rm ChM,z}\) as cleanly as the continuous orbital variables:
the GSE class alone spans nearly the full range of the index, with NGC~2808 at
one extreme. 

This result extends our previous comparison of MP properties with cluster origin
\citep{lardo26}, which found little evidence that global MP diagnostics, such as
the fraction of 1P stars \((f_{\rm 1P})\) and the helium abundance range
\((\Delta Y)\), depend strongly on in-situ versus accreted origin once mass and
metallicity are controlled for. The present result suggests that the corrected
ChM extension is not fixed by progenitor label alone. Instead, it appears to
trace a more continuous environmental/orbital dimension, of which dynamical
origin is only one component. A host-galaxy imprint on MP-related photometric properties was first noted by
\citet{dalessandro12}, who showed that integrated \((FUV-V)\) colours of GCs
become progressively bluer moving from the Sagittarius dwarf to the Milky Way,
M\,31, and M\,87, suggesting that more massive hosts produce GCs with stronger
UV excess --- a trend qualitatively aligned with the cluster-mass and
orbital-confinement scaling identified here.

\subsection{Interpretation: birth environment versus survival bias}
\label{sec:environment_discussion}

Two interpretations of the orbital signal must be separated. Direct causality
can be excluded: the present-day orbit cannot have produced the light-element
abundance patterns, which were established within the first few \(10^{8}\) yr of
the cluster's life, long before the present orbit was set. The orbit can only
trace the conditions under which the cluster formed, or select 
which clusters survived. We test the selective (survival-bias) interpretation in
detail before turning to the fossil interpretation.

\paragraph{Is the orbital trend a survival bias?}
Clusters on confined, inner orbits experience stronger disc shocking, tidal
stripping, and evaporation. If clusters with intrinsically weak MP signatures
were preferentially destroyed there, the surviving inner population could appear biased toward clusters with
larger ChM extensions, either because these were initially more massive or
because their enriched populations were more centrally bound. We test this at three levels.

First, the index does not track dynamical erosion. We compared
\(S_{\rm ChM,z}\) with fractional mass loss, dissolution time, half-mass
relaxation time, present-day mass-function slope, and current mass; none shows a
significant correlation after correction for the number of quantities tested.
The \(S_{\rm ChM,z}\) index therefore does not
behave as a tracer of how dynamically eroded a cluster is today. This weakens
the simplest survival-bias interpretation, in which the ChM extension would be
expected to scale directly with cumulative mass loss or relaxation-driven
evolution.

Second, the orbital trend is not absorbed by survival proxies. Within the
low-reddening subsample, where the signal is cleanly measured
(Appendix~\ref{app:reddening_robustness}), the \(S_{\rm ChM,z}\)--orbit
anti-correlations remain after controlling jointly for initial mass and
metallicity (e.g. \(z_{\max}\): \(\rho_{\rm S}=-0.70\), \(p=0.004\)), and
likewise when representative dynamical-evolution indicators are added as
controls. Thus, the orbital dependence is not simply a re-expression of the
fact that more confined clusters are, on average, more dynamically processed.

Third, the geometry of the trend is inconsistent with simple tidal stripping.
Restricting to the low-reddening subsample (where the reddening-sensitive
\(L_\perp^{2P}\) can be set aside; Appendix~\ref{app:reddening_robustness}),
the orbital signal is carried by the mean 2P offset
\(\Delta_\parallel^{2P}\) (\(\rho_{\rm S}=-0.63\), \(p=0.012\)) and the
 \(L_\parallel^{2P}\) (\(\rho_{\rm S}=-0.59\),
\(p=0.022\)). Preferential stripping of the least-enriched 2P stars would increase
\(\Delta_\parallel^{2P}\) while reducing \(L_\parallel^{2P}\), producing a
truncation-like ChM geometry in which the enriched sequence becomes more
displaced but less extended. Instead, on more confined orbits,
\(\Delta_\parallel^{2P}\) and \(L_\parallel^{2P}\) increase together. This
coherently longer and more displaced enriched sequence is therefore more
consistent with a fully developed 2P than with one truncated by stellar loss.

Taken together, these tests argue against survival bias as the sole origin of
the observed relation. Survival effects may still contribute, since clusters
on more confined orbits show signs of stronger tidal processing in their
mass-function slopes and relaxation times. Yet the two key expectations of a
survival-driven trend are not observed: \(S_{\rm ChM,z}\) does not correlate
directly with erosion indicators, and the ChM geometry is not truncation-like.
The orbital dependence of \(S_{\rm ChM,z}\) therefore appears to reflect an
environmental imprint beyond mass, metallicity, and the available measures of
dynamical evolution.

\paragraph{A fossil imprint of the birth environment.}
If the orbital trend is not produced primarily by present-day dynamical erosion,
the remaining interpretation is that it preserves a fossil imprint of the birth
or early evolutionary environment. The local density required to form a massive
bound cluster may be similar across galactic environments, but the larger-scale
boundary conditions need not be: the gas reservoir, external confining pressure, tidal
field, and supply of pristine gas available for dilution can all vary with host
environment and orbital confinement
\citep[e.g.][]{kruijssen12,kruijssen15,reinacampos17,pfeffer18,adamo20}. 

In this picture, clusters born in more confined, gas-rich environments could
experience a longer or more efficient enrichment--dilution cycle. This does not
require the local density of the proto-cluster itself to differ strongly.
Rather, the surrounding medium may regulate how long processed ejecta and
diluting gas remain available before being expelled or stripped. The observed
trend would then reflect the environment controlling the duration and
efficiency of 2P formation, rather than the local density of the initial bound
cluster alone. This interpretation is consistent with two empirical features of our analysis.
First, mass is not sufficient to explain the trend: at fixed mass and
metallicity, orbital confinement still modulates \(S_{\rm ChM,z}\)
(Sect.~\ref{sec:drivers}). Second, continuous orbital variables organise the
index more cleanly than discrete progenitor labels, suggesting that the relevant
quantity is not simply accretion origin, but the degree of orbital confinement
or the environmental conditions associated with it.

\paragraph{}

The existence of MPs may therefore be nearly universal among massive Galactic
GCs, while the detailed morphology of the enriched sequence retains a subtler
environmental memory. In this sense, \(S_{\rm ChM,z}\) is sensitive not only to
whether MPs are present, but also to how extended and chemically diverse the
enriched population became. A caveat concerns the chemically complex Type~II clusters. Systems with
large anomalous fractions, which could bias the ChM verticalisation and
metallicity correction, were excluded; when Type~II clusters were retained, the
CNO-enhanced component was explicitly removed from the analysis (Appendix~\ref{app:phot_assignment}). The remaining
Type~II clusters are therefore unlikely to drive the trend through a separate
anomalous sequence. Instead, they may retain additional information about more
complex formation channels not captured by the standard light-element sequence
alone \citep{marino19,pfeffer21}. A larger sample, with Type~I and Type~II clusters treated separately, will be
needed to determine whether Type~II morphology provides a complementary part of
the \(S_{\rm ChM,z}\)--environment signal.

\section{Summary and conclusions}
\label{sec:summary}

We have developed a physically corrected ChM framework for a sample of 23
Galactic GCs. The method removes the dominant metallicity- and
reddening-dependent photometric response from the raw HST ChM, while
preserving the residual cluster-to-cluster morphology as the astrophysical
signal of interest. For each cluster we measure the 2P sequence in a
local enrichment-aligned frame through three observables --- its median
displacement from the 1P locus \(\Delta_\parallel^{2P}\), its extent along the
enrichment direction \(L_\parallel^{2P}\), and its perpendicular extent
\(L_\perp^{2P}\) --- and combine them into a standardised photometric extension
index, \(S_{\rm ChM,z}\). Our main conclusions are as follows.

\begin{enumerate}

\item The raw ChM is not a universal coordinate system. The 1P and 2P loci shift
systematically with metallicity, and the apparent 1P--2P separation is partly
amplified by the metallicity-dependent response of the UV filters, so raw ChM
separations cannot be compared directly across clusters. Correcting for this
response yields a common, physically meaningful frame in which the three 2P
observables vary coherently along a single cluster-to-cluster amplitude axis;
their first principal component defines \(S_{\rm ChM,z}\).

\item APOGEE spectroscopy validates the chemical meaning of the corrected
morphology at the population level. \(S_{\rm ChM,z}\) correlates not with any
single abundance ratio but with the multivariate chemical width of the 2P
population along a high-temperature proton-capture axis (dominated by Mg and O
depletion and Al enhancement). It is therefore a tracer of the internal chemical
diversity of the enriched population, not a proxy for the mean 1P--2P
abundance offset.

\item The extent of the enriched morphology is not regulated by cluster mass
alone. Initial mass is positively but only marginally associated with
\(S_{\rm ChM,z}\) in this sample, whose mass coverage is narrow and biased toward
massive systems. Hence, the established mass dependence of the MP phenomenon is not
contradicted, merely weakly sampled. The strongest correlations are instead with
a covariant family of orbital quantities (i.e.  \(z_{\max}\), \(J_z\), orbital
energy, apocentre, and Galactocentric radius) such that clusters on more
confined, inner orbits have more extended corrected ChM morphologies. 

\item This orbital signal is not an artefact of foreground reddening. Although
orbit and reddening are degenerate across the full sample, the correlations
survive and strengthen in the low-reddening subsample where the two are
decoupled. Within that clean
subsample, mass and orbital confinement emerge as two independent
contributions to the ChM extension, with the orbital contribution the stronger
over the range our sample probes.

\item The orbital trend is best read as a fossil imprint of the birth
environment rather than a survival artefact. The present orbit cannot have
caused the early-established abundance pattern; it can only trace formation
conditions or select survivors. Dynamical-erosion proxies do not reproduce the
trend, and its geometry is inconsistent with tidal stripping. 
Discrete progenitor labels vary continuously with this same orbital structure and add
no information beyond it. A residual survival contribution cannot be excluded,
and forward modelling will be needed to fully separate birth-environment effects
from orbit-dependent survival.

\end{enumerate}

The corrected ChM thus provides a chemically validated measure of the internal
diversity of the enriched population, and the novel result of this work is that
this diversity carries an orbital/environmental imprint beyond cluster mass and
metallicity. Read as a fossil record of the birth environment, this implies that
MP formation scenarios must reproduce not only the internal polluter budget set
by the cluster potential, but also the dependence of the 2P morphology on
orbital confinement.

Future work should extend the chemical validation to a larger homogeneous
spectroscopic sample, particularly at the metal-poor and metal-rich extremes;
apply the corrected framework to larger \textit{HST} and JWST photometric
samples; and connect present-day orbital confinement to birth environment using
cosmological simulations of Galactic GC assembly. Together these will clarify
whether \(S_{\rm ChM,z}\) primarily records natal gas conditions, early retention
and dilution, subsequent tidal evolution, or a combination.

\section*{Data availability}
The code and procedures used to construct the chromosome maps and reproduce the
photometric cleaning, binary-candidate removal, and multiple-population assignment
presented in this work are publicly available on Zenodo at
\url{https://doi.org/10.5281/zenodo.21487109}.

\begin{acknowledgements}
This research was supported by the International Space Science Institute (ISSI) in Bern, through ISSI International Team project 25-636 {\em The Origin of Multiple Populations in Globular Clusters}.      
\end{acknowledgements}

\bibliographystyle{aa}  
\bibliography{biblio}

\appendix

\section{Photometric cleaning and population assignment}

\subsection{Cleaning of the chromosome maps}
\label{app:phot_cleaning}

The construction of ChM morphology diagnostics requires a homogeneous RGB
selection and a reproducible separation between primordial and enriched
populations. We therefore applied the same cleaning and classification
procedure to all clusters before measuring the ChM observables used in the main
analysis. This appendix describes the two-stage pipeline used to remove field
contamination and mitigate the impact of unresolved binaries, both of which can
broaden the observed ChM sequences \citep[e.g.][]{marino3201,martins2020}. 

\paragraph{Field-star separation}

\begin{figure*}
    \centering
    \includegraphics[width=1\linewidth]{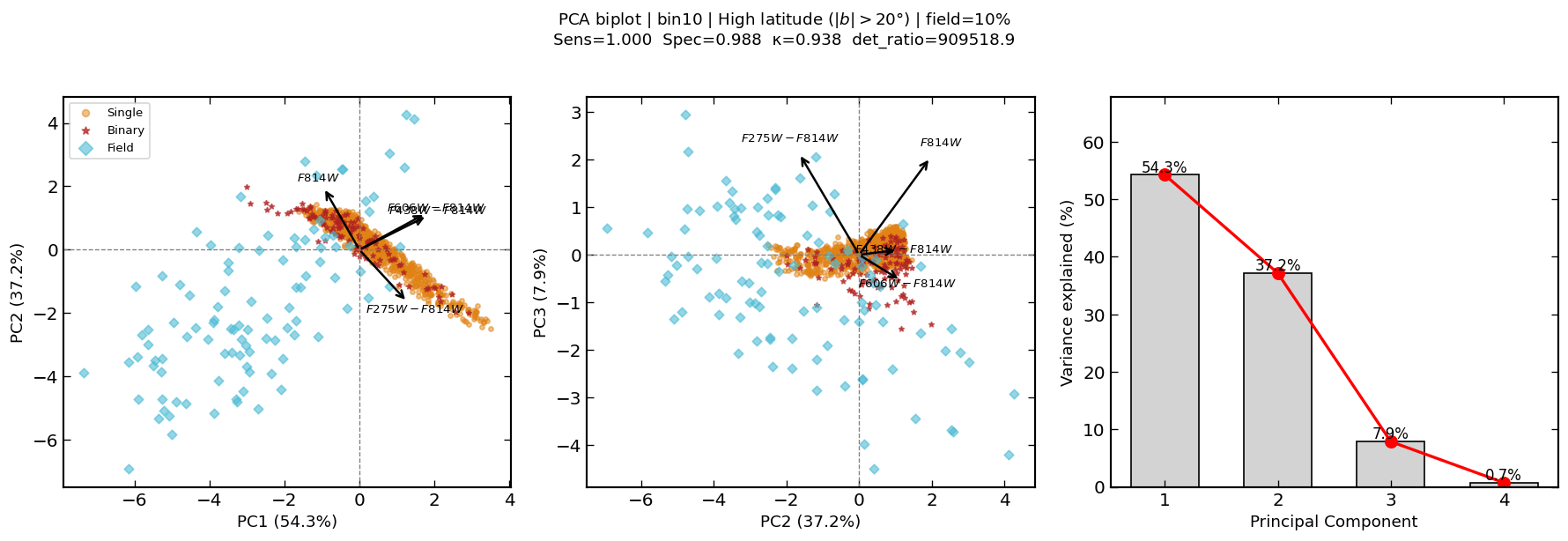}
    \caption{PCA visualisation of the four-dimensional colour--magnitude space
\((F275W-F814W, F438W-F814W, F606W-F814W, F814W)\) used for field-star
separation, shown for the synthetic high-latitude environment with a binary
fraction of 10\% and a field-contamination fraction of 10\%.
{\em Left:} PC1--PC2 score plot. {\em Centre:} PC2--PC3 score plot.
{\em Right:} scree plot showing the fraction of variance explained by each
principal component. Orange circles are single cluster stars, red stars are
unresolved binaries, and cyan diamonds are injected field stars. Black arrows
show the PCA loading vectors. The first two principal components capture most
of the variance and separate the compact cluster locus from the broader field
distribution. For this realisation, the field-rejection performance is high,
with sensitivity \(=1.000\), specificity \(=0.988\), Cohen's
\(\kappa=0.938\). The large determinant ratio
\({\rm det}_{\rm ratio}=9.10\times10^{5}\) quantifies the much greater
compactness of the recovered cluster locus relative to the field component.
}
    \label{fig:highlat}
\end{figure*}

Field stars are removed using a two-component GMM fitted in the four-dimensional colour–
magnitude space: \[
(F275W-F814W,\; F438W-F814W,\; F606W-F814W,\; F814W).
\]
Cluster RGB stars occupy a compact locus in this space, while field stars form a broader distribution. The cluster component is identified as the one with the smallest covariance determinant, providing a simple and re-producible criterion. To visualise the structure of this space we also perform a PCA  of the same
observables. As shown in Fig.~\ref{fig:highlat}, the first two principal components account for \( \sim 90\%\) of the variance and cleanly separate the compact cluster locus from the more diffuse field population. 
Validation on synthetic catalogues sampling a range of Galactic environments
and field-contamination fractions (5--50\%) shows that the GMM classification
retains 98--99\% of cluster members while achieving near-complete recovery of
field stars, with Cohen's \(\kappa \sim 0.9\)--0.99. The latter quantifies the
agreement between recovered and true cluster/field labels after correcting for
chance agreement, with \(\kappa=1\) indicating perfect classification and
\(\kappa=0\) random agreement. 

\paragraph{Binary-star mitigation}

Candidate unresolved binaries and other broadening contaminants (i.e. photometric blends) were then
mitigated with a second GMM applied in the verticalised
pseudo-colour space (Sect.~\ref{sec:gcs}):
\begin{equation}
\begin{split}
\mathbf{x}_{\rm bin} =
\bigl(&\Delta(F275W-F814W),\\
&\Delta C_{F275W,F336W,F438W},\\
&\Delta C_{F336W,F438W,F814W}\bigr).
\end{split}
\end{equation}
The choice of these axes is supported by a PCA
of the verticalised colours, which demonstrates that the dominant axes of variation are driven by light-element abundance differences -- the physical signature of MPs --
whereas binary stars scatter nearly uniformly around the normalized colour locus, orthogonally to the main population sequence. The used pseudo-colours are therefore efficient discriminators of genuine population spread from binary-induced broadening.

The most diffuse component was flagged as a candidate-binary or broadening
component and removed from the ChM analysis. Synthetic tests with different
field-contamination and binary fractions show that this procedure preserves
the compact RGB locus while removing the dominant asymmetric broadening
component. The performance of the binary-removal
stage was evaluated for injected binary fractions of 10\%  and 30\%, evaluated across three Galactic environments and field
fractions from 5\% to 50\%. In both binary-fraction cases, the specificity is
uniformly high, \(0.99 -1.00\), meaning that genuine single-star members are
rarely misclassified as binaries. The sensitivity is more modest, remaining at
\(\sim0.55\) and \(\sim0.61\) for the 10\% and 30\% binary cases,
respectively, for field fractions up to 30\%. Cohen's \(\kappa\) values of
\(\sim0.68\) and \(\sim0.66\) indicate good agreement with the known input
labels after correcting for chance agreement. Thus, this step is effective as
a conservative mitigation of binary-induced broadening, although it is not a
complete binary census.

Performance degrades only in the most challenging synthetic configuration,
with 50\% field contamination. In this case,
the sensitivity drops to 0.54 and 0.31 for the 10\% and 30\% binary cases,
respectively, and \( \kappa \) falls to 0.37 in the latter case. This reflects
the increased difficulty of separating binary-displaced stars from a heavily
contaminated field and represents the stress-test limit of the method. In real
data, the efficiency of this step will also depend on the binary mass-ratio
distribution, photometric depth, crowding, and residual differential reddening.

\paragraph{Limitations and applicability of the cleaning procedure}
The synthetic validation confirms the robustness of the cleaning
procedure under controlled conditions, but does not capture all
sources of observational complexity. Differential reddening, spatially varying completeness, and non-Gaussian field populations may introduce additional scatter not accounted for in the tests
above. The GMM-based approach should therefore be regarded as an empirical and homogeneous selection method rather than a definitive membership classification, with its primary strength
lying in its consistency across the full cluster sample. Any residual contamination would need to correlate systematically with cluster properties in order to bias the comparative results presented in the main text. However, we find no evidence for such a correlation.

\subsection{Population assignment}
\label{app:phot_assignment}
The final population assignment was performed with a Gaussian mixture model in
a five-dimensional verticalised colour--pseudo-colour space,
\begin{equation}
\begin{split}
\mathbf{x}_{\rm pop} =
\bigl(&\Delta(F275W-F814W),\;
\Delta(F438W-F814W),\\
&\Delta(F606W-F814W),\;
\Delta C_{F275W,F336W,F438W},\\
&\Delta C_{F336W,F438W,F814W}\bigr).
\end{split}
\end{equation}
All dimensions were standardised to zero mean and unit variance before fitting.

\begin{figure*}
\sidecaption
\includegraphics[width=12cm]{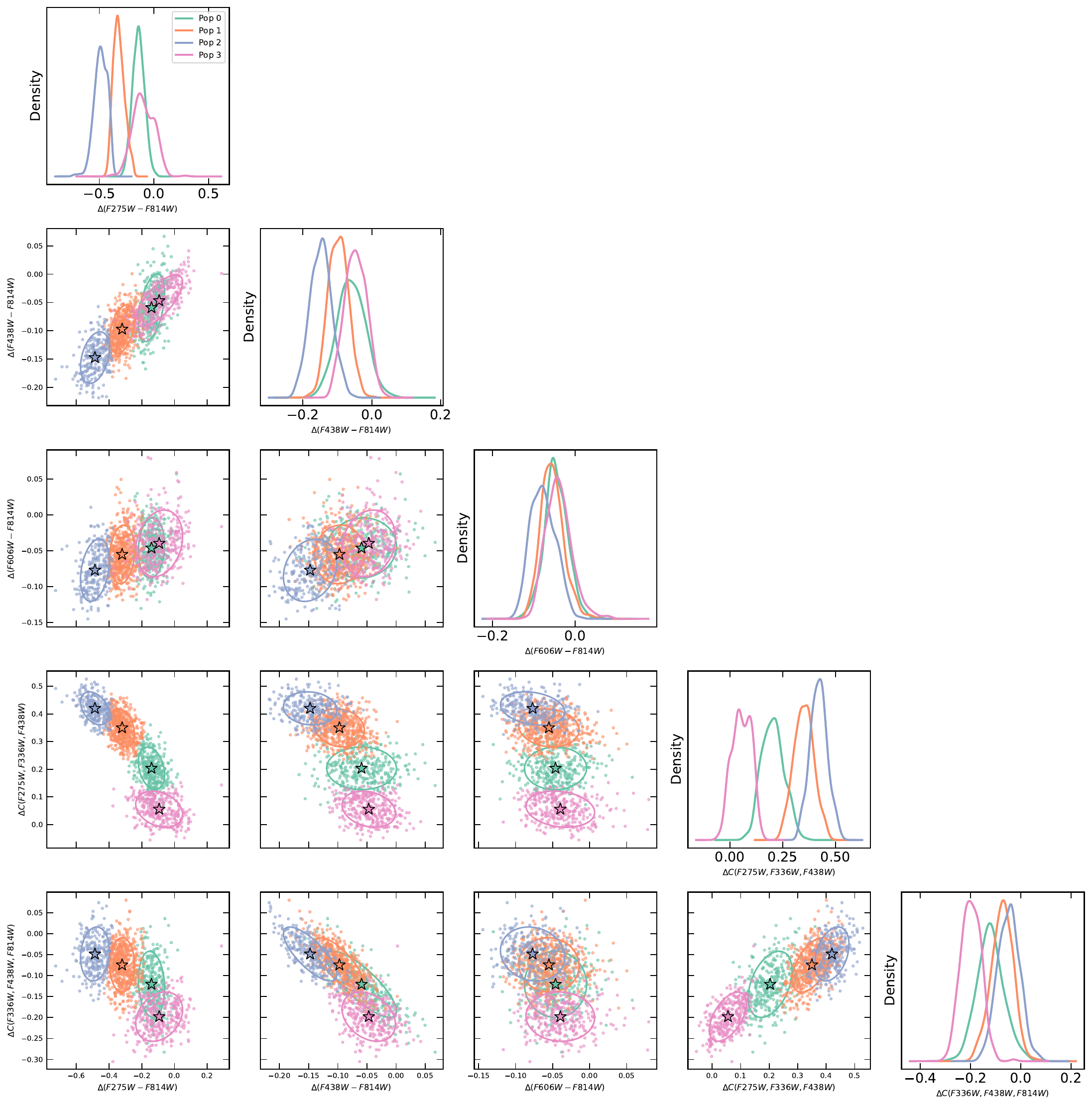}
\caption{
Pseudo-colour corner plot for NGC~2808 after field-star and candidate-binary
removal. Diagonal panels show the marginal distributions of the five
pseudo-colour indices, while off-diagonal panels show the corresponding
bivariate projections colour-coded by the GMM population assignment. The main
analysis uses only the coarse 1P/2P partition: the 1P component defines the
primordial reference locus, and the remaining components define the enriched
sequence whose morphology is measured in the corrected ChM frame.
}
\label{fig:corner}
\end{figure*}

The choice of pseudo-colour axes is physically motivated by
the spectral response of each HST filter to the light-element
and helium abundance variations that define MPs. The \(F275W\) bandpass straddles the OH band near
3080~\( \AA\), making it sensitive to oxygen depletions; \(F336W\) samples the NH band at \( \sim 3360 ~\AA\) and CN features; \(F438W\) captures the CH G-band and CN bands that respond in the opposite sense to N enhancement \citep[e.g.;][]{piotto15,salaris20}. The 
pseudo-colour 
\[ \Delta C_{F275W,F336W,F438W} \equiv  \Delta [(F275W-F336W)-(F336W-F438W)] 
\] 
amplifies the N-abundance signature while partially cancelling temperature- and metallicity-
driven continuum slopes, since the three filters share a common
\(T_{\rm eff} \) dependence. In practice, a nitrogen-rich (2P) star drives
 \( \Delta C_{F275W,F336W,F438W}\) to more positive values, producing
the pronounced vertical separation visible in the chromosome
map \citep{milone18}. The pseudo-colour  
\[
\Delta C_{F336W,F438W,F814W} \equiv \Delta [(F336W-F438W)-(F438W-F814W)] 
\]
extends the
baseline to the optical. Because \(F814W\) lies far red-ward of the
UV molecular bands, this index is sensitive both to N-cycle
chemistry and to helium abundance, since a helium-enhanced
star is hotter and bluer along the RGB. The two pseudo-colour
indices share \(F336W\) and \(F438W\), so variations in those bands
produce correlated shifts in both axes simultaneously (Fig.~\ref{fig:corner}).

The remaining broad-baseline colours help reduce this degeneracy. The colour
\(\Delta(F275W-F814W)\) is primarily sensitive to the temperature displacement
of the RGB, and therefore to helium-related shifts, with additional sensitivity
to UV line blanketing  \citep[e.g.][]{lardo18,milone18}. The optical colours
\(\Delta(F438W-F814W)\) and \(\Delta(F606W-F814W)\) are less directly sensitive
to the UV molecular bands but respond to the RGB temperature locus and to
residual broadening in the optical colours \citep[e.g.;][]{sbordone11}.
Together, the five axes span a space in which the temperature-sensitive
broad-baseline colours and the N-sensitive UV pseudo-colours provide
complementary information for separating the compact 1P locus from the
enriched sequence.

We model this joint distribution with a full-covariance GMM, allowing each
component to have its own unconstrained covariance matrix and thereby capturing
the tilted morphology of the ChM sequences. The number of components \(k\) is
varied from 1 to 5, with the optimal value selected by the minimum BIC. Field
stars and candidate binaries, identified in the previous cleaning stages
(Appendix~\ref{app:phot_cleaning}), are removed before this fit. The component
whose centroid lies closest to the compact reference locus around the ChM
origin is identified as the 1P component, while the remaining components are
grouped into a single coarse 2P class. This 1P/2P partition is the only
population label used in the main analysis. It is deliberately agnostic about
finer sub-populations and is used only to define the 1P reference locus and the
morphology of the enriched sequence.

The example corner plot for NGC~2808 (Fig.~\ref{fig:corner}) illustrates the
behaviour of this five-dimensional classification. The strongest separation
between 1P and 2P stars occurs along
\(\Delta C_{F275W,F336W,F438W}\) and \(\Delta(F275W-F814W)\), while
\(\Delta C_{F336W,F438W,F814W}\) adds discriminating power mainly within the
enriched sequence. The tilted ellipses show why a full-covariance model is
needed: the relevant ChM axes are correlated rather than independent.

The colour--colour structure of the 1P component also motivates our choice not
to use the 1P width as a primary measure of MP strength. In several projections
of Fig.~\ref{fig:corner}, the 1P-like component is not simply broadened
isotropically but shows a distinct covariance direction relative to the
enriched components. This behaviour is consistent with residual structure
within the nominal 1P population, which may reflect small pristine-abundance
variations, residual photometric systematics, or a combination of both. We therefore use the 1P locus only as the reference against which the enriched
sequence is measured, and base the ChM-extension index entirely on the
morphology of the 2P (Sect.~\ref{sec:chm_strength}).

Clusters with known anomalous sub-populations were treated conservatively. In
Type~II or chemically complex clusters where the CNO-enhanced population
constitutes a substantial fraction of the RGB population, such as NGC~5286 and
M~22, the anomalous sequence can shift the reference locus identified as 1P and
alter the inferred 2P morphology. These systems were therefore excluded from
the analysis. Clusters in which the anomalous component is only a minority population were
retained, but the CNO-enhanced stars were explicitly removed before measuring
the ChM observables. This prevents the anomalous sequence from affecting the
cluster-level 1P locus, 2P centroid, or 2P extent at the precision relevant
here. This treatment applies, for example, to NGC~1851, M~15, and NGC~362.

\section{APOGEE sample selection and chemical classification}
\label{app:chem_classification}

We use the homogeneous APOGEE value-added catalogue of \citet{schiavon24} to
validate the photometric ChM observables against spectroscopic abundances
(Sect.~\ref{sec:chemical_validation}). The spectroscopic population
classification is performed entirely in abundance space, before any comparison
with the ChM observables, ensuring that the chemical validation is independent
of the photometric ChM coordinates.

\paragraph{Quality cuts and abundance space}
\label{app:apogee_cuts}

Stars are retained if they have high cluster-membership probability
\((P_{\rm mem}>0.80)\), spectral signal-to-noise ratio
\({\rm S/N}>50\), and \(\log g<3.6\), selecting probable RGB cluster members.
We require valid per-element
abundance uncertainties in the range \(0.03\)--\(0.50\) dex for all elements
used in the chemical classification.

The chemical classification is then performed in the five-dimensional
light-element abundance space 
\(
\bigl([\mathrm{C/Fe}],[\mathrm{N/Fe}],[\mathrm{O/Fe}],
[\mathrm{Mg/Fe}],[\mathrm{Al/Fe}]\bigr),
\)
which captures the main proton-capture abundance variations associated with
MPs. The ChM observables are introduced only after the chemical
classification, and only at the cluster level, to test whether the corrected
photometric morphology traces the spectroscopic chemical diversity of the
2P.

\begin{figure*}
\sidecaption
\includegraphics[width=12cm]{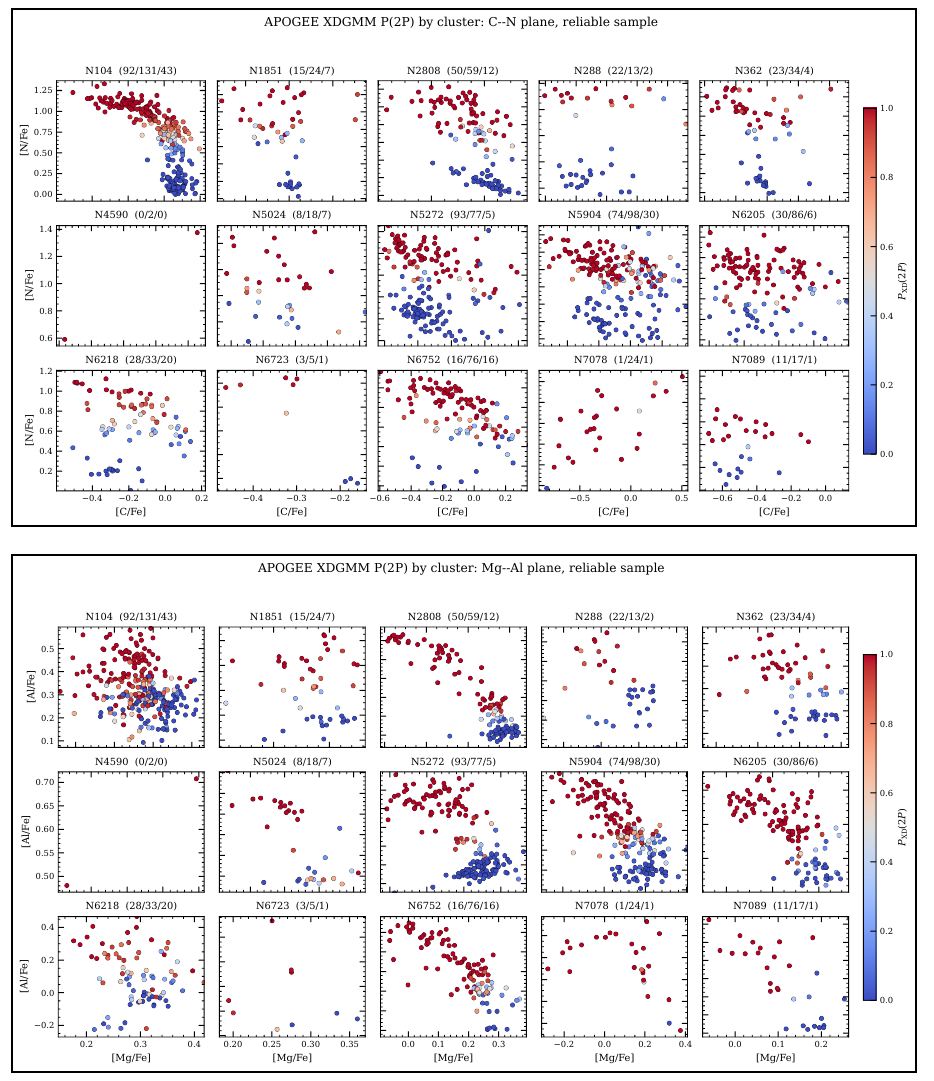}
\caption{
Example APOGEE chemical classification from the XDGMM analysis in the
C--N and Mg-Al diagrams. Each panel shows one cluster, with points colour-coded by the
posterior probability of belonging to the enriched population,
\(P_{\rm XD}(2P)\). Blue points correspond to chemically 1P-like stars,
red points to chemically 2P-like stars, and intermediate colours indicate
ambiguous objects. The numbers in parentheses give the number of XDGMM
1P, 2P, and ambiguous stars, respectively. The classification is performed
in the full five-dimensional abundance space
\([\mathrm{C/Fe}], [\mathrm{N/Fe}], [\mathrm{O/Fe}],
[\mathrm{Mg/Fe}], [\mathrm{Al/Fe}]\), and is independent of the
photometric ChM coordinates. We include both reliable validation clusters
and examples of unreliable cases; for instance, NGC~4590 has only two
APOGEE stars in the selected sample, and no chemically classified 1P stars,
therefore it is excluded from the cluster-level validation.}
\label{fig:apogee_xdgmm_cn}
\end{figure*}

\paragraph{XDGMM chemical classification}
\label{app:xdgmm_classification}

For each cluster, we model the distribution of APOGEE stars in the
five-dimensional abundance space with an extreme-deconvolution Gaussian
mixture model \citep[XDGMM;][]{Bovy2011}. This method accounts for the
individual abundance uncertainties and avoids assigning population labels from
a single abundance ratio or two-dimensional chemical projection.

The chemically 1P-like component is identified as the component with lower
proton-capture enrichment, while the chemically 2P-like component is identified
as the component enhanced in the processed material. Operationally, this means
that the 2P component occupies the direction of enhanced N and Al and depleted
C, O, and, where measured, Mg. Stars with high posterior probability \((P>0.80)\) are
assigned to the corresponding 1P or 2P class, while stars with intermediate
posterior probabilities are labelled as ambiguous.

\paragraph{Reliable cluster-level validation sample}
\label{app:reliability}

The purpose of the APOGEE comparison is not to classify individual stars for
its own sake, but to derive robust cluster-level chemical quantities: the
2P chemical width, the mean 2P--1P abundance offset, and the extent of the
chemically enriched population (Sect.~\ref{sec:chemical_validation}). We therefore define a reliable validation sample using minimum-number
requirements on the XDGMM-classified populations:
\[
N_{\rm XD,1P} \geq 10, \qquad
N_{\rm XD,2P} \geq 5, \qquad
f_{\rm XD,1P} \geq 0.03 .
\]
Here, \(N_{\rm XD,1P}\) and \(N_{\rm XD,2P}\) are the numbers of APOGEE stars
classified by the XDGMM as first- and second-population members, respectively,
while \(f_{\rm XD,1P}\) is the corresponding fraction of XDGMM-classified 1P
stars in the cluster. Clusters that fail one or more of these criteria are
retained in diagnostic figures but excluded from the quantitative validation
tests. This distinction is important because some systems have APOGEE coverage
but too few stars in one of the two chemical components to define stable
cluster-level statistics. For example, NGC~4590 has only two selected APOGEE stars, both classified as
2P-like, and no chemically classified 1P reference population. It is therefore
shown in Fig.~\ref{fig:apogee_xdgmm_cn} as an illustrative unreliable case but excluded from the validation
analysis. The final reliable APOGEE validation sample contains 11 clusters.
Figure~\ref{fig:apogee_xdgmm_cn} shows the chemical classification 
from the XDGMM analysis in the C--N and Mg-Al diagrams.

\paragraph{Ambiguous stars}
\label{app:ambiguous}

Stars with intermediate XDGMM posterior probability are labelled ambiguous and
excluded from the cluster-level 1P and 2P summary statistics. This conservative
choice prevents chemically intermediate objects or low-precision abundance
measurements from artificially broadening the inferred 2P distribution.

Ambiguous stars can arise for two reasons. They may be genuinely intermediate,
reflecting formation out of gas with different degrees of dilution between
pristine material and processed ejecta. Alternatively, they may result from
measurement uncertainties that blur the separation between the chemical
components, especially in clusters where the intrinsic 1P--2P contrast is
small. In both cases, excluding ambiguous stars makes the APOGEE validation
conservative: the test is whether the robustly classified enriched population
has a multivariate chemical width that corresponds to the corrected ChM
morphology.

The probability maps shown in Fig.~\ref{fig:apogee_xdgmm_cn} show  this
classification scheme. In several clusters, the posterior probabilities vary
smoothly between the 1P-like and 2P-like components, denoting  that ambiguous
stars typically occupy an intermediate chemical space rather than forming an
independent third population.

\section{Robustness of the full-sample driver analysis}
\label{app:robustness}

The orbital correlations reported in Sect.~\ref{sec:drivers} could in principle
be confounded by foreground reddening, baseline chemical differences, or
formation epoch. We address each possibility in turn.

\begin{figure}[!ht]
\centering
\includegraphics[width=0.9\columnwidth]{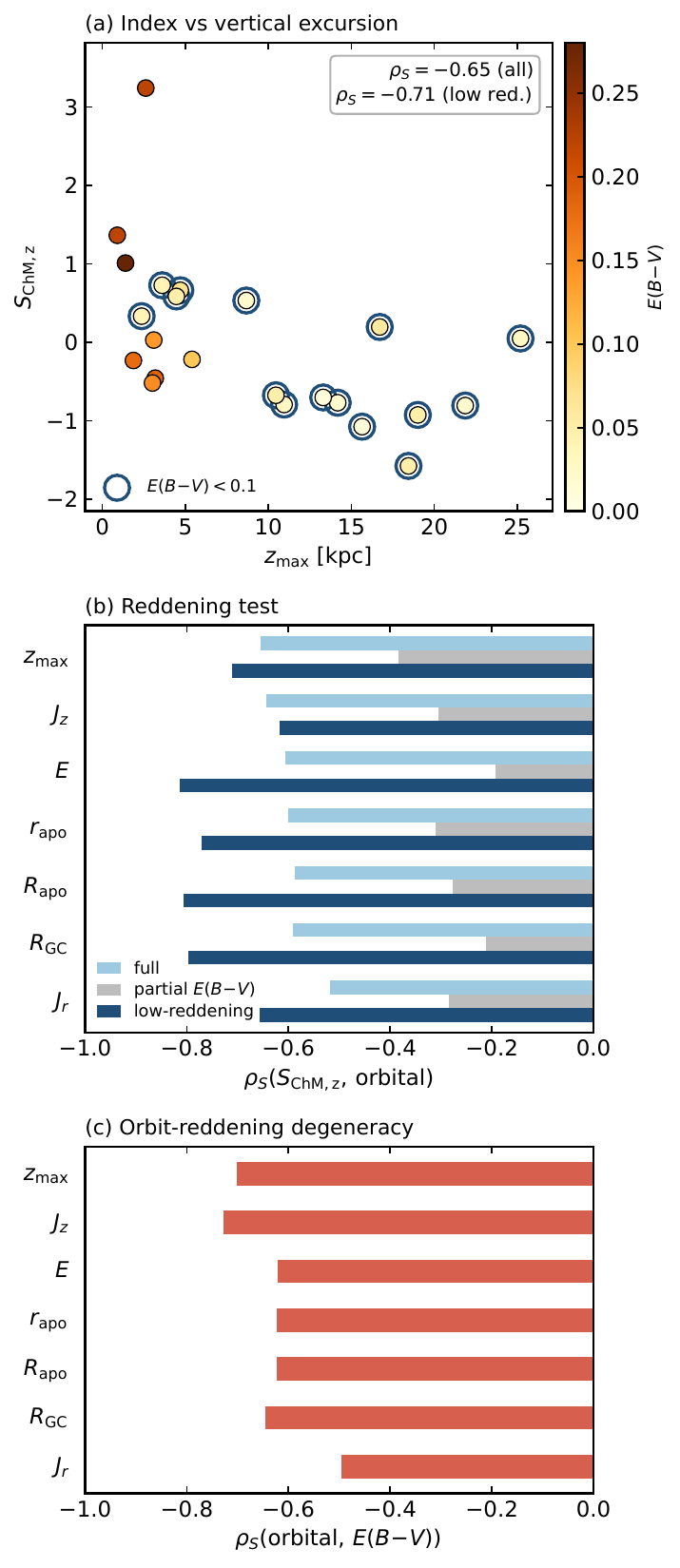}
\caption{Orbital correlations of the ChM strength index and the reddening test.
\textit{(a)} \(S_{\rm ChM,z}\) versus 
\(z_{\max}\), with points coloured by foreground reddening \(E(B-V)\); rings mark
the low-reddening subsample (\(E(B-V)<0.1\)). Spearman correlation for the full sample and the low-reddening subsample are annotated. \textit{(b)} Spearman correlation of \(S_{\rm ChM,z}\) with each orbital quantity
in the full sample (light), after partialling out \(E(B-V)\) (grey), and in the
low-reddening subsample (dark). 
\textit{(c)} Spearman correlation of each orbital quantity with \(E(B-V)\).}
\label{fig:orbital_summary}
\end{figure}

\subsection{Foreground reddening}
\label{app:reddening_robustness}
The orbital-confinement quantities are correlated not only with one another but
with foreground reddening: across the full sample, \(z_{\max}\), \(J_z\),
orbital energy, apocentre, and Galactocentric radius all correlate with
\(E(B-V)\) at \(|\rho_{\rm S}|\simeq0.6\)--\(0.7\)
(Fig.~\ref{fig:orbital_summary}c). This is expected, because dust is
concentrated in the inner Galactic disc where dynamically confined clusters
spend most of their time, and it raises the concern that the
\(S_{\rm ChM,z}\)--orbit relation is in part a residual reddening effect. The
concern is sharpened by the fact that the metallicity- and reddening-dependent
correction applied to the ChM (Sect.~\ref{sec:chm_strength}) removes only the
mean foreground term: it shifts each cluster's map to a common reference
\(E(B-V)\), but cannot remove differential reddening within a cluster,
which broadens the enriched sequence and therefore preferentially affects the
perpendicular width \(L_\perp^{2P}\).

Because orbit and reddening are this collinear, a
partial correlation cannot separate them: partialling out \(E(B-V)\) removes the
variance shared by the two predictors --- which contains the genuine orbital
signal as well as any reddening contribution --- and therefore attenuates a real
effect rather than isolating a spurious one. Indeed, when \(E(B-V)\) is
partialled out the orbital correlations all weaken (Fig.~\ref{fig:orbital_summary}b,
grey bars), but this attenuation is not by itself diagnostic of the signal's
origin.

We therefore use a more decisive test: restricting the sample to the 15 clusters
with low foreground reddening, \(E(B-V)<0.1\). Within this subsample orbit and
reddening are effectively decoupled --- \(z_{\max}\) and \(E(B-V)\) are
uncorrelated (\(\rho_{\rm S}=-0.17\), \(p=0.54\)) --- so reddening can no longer
generate an orbital correlation. The orbital signal is nonetheless fully
preserved, and for several quantities strengthens
(Fig.~\ref{fig:orbital_summary}b, dark blue bars):
\[
\begin{aligned}
\rho_{\rm S}(S_{\rm ChM,z},\,z_{\max})   &= -0.71 \;\;(p=0.003),\\
\rho_{\rm S}(S_{\rm ChM,z},\,r_{\rm apo}) &= -0.77 \;\;(p=0.001),\\
\rho_{\rm S}(S_{\rm ChM,z},\,R_{\rm GC})  &= -0.80 \;\;(p<0.001).
\end{aligned}
\]
Within this clean subsample the orbital correlations also survive joint control
for initial mass and metallicity (e.g. \(z_{\max}\): \(\rho_{\rm S}=-0.70\),
\(p=0.004\); \(R_{\rm GC}\): \(-0.85\), \(p<0.001\)). We conclude that the
orbital-confinement signal is not produced by foreground reddening: the collapse
of the partial correlations in the full sample reflects the orbit--dust
collinearity, not the absence of an orbital effect, and the signal is recovered where
the two are physically disentangled.

The same subsample clarifies the role of mass. With orbit and reddening
decoupled, initial mass shows an independent positive correlation with
\(S_{\rm ChM,z}\) (\(\rho_{\rm S}=+0.53\), \(p=0.04\); rising to \(+0.64\) when
\(z_{\max}\) is partialled out), and mass and \(z_{\max}\) are themselves
uncorrelated within the subsample (\(\rho_{\rm S}=-0.09\)).

\subsection{Controlling for chemistry and age.}
\label{app:age}

Two further checks address whether the orbital signal could instead reflect
baseline chemical differences or formation epoch. The first concerns
\(\alpha\)-enhancement. Clusters associated with different Galactic components
or accreted progenitors may have formed from gas with different
\([\alpha/\mathrm{Fe}]\) ratios. If these baseline abundance differences were
correlated with orbit, the \(S_{\rm ChM,z}\)--orbit relation could be a disguised
chemical-tagging effect rather than a genuine dependence of the MP morphology
on environment. We therefore tested this possibility in the APOGEE overlap
sample, where homogeneous first-population abundances are available for
11 clusters. We used the median \([\mathrm{Mg/Fe}]_{\rm 1P}\) of the
XDGMM-classified 1P stars as a proxy for the baseline \(\alpha\)-enhancement,
noting that 2P Mg can be altered by Mg--Al processing and is therefore not a
clean tracer of the initial composition. Within this subset,
\([\mathrm{Mg/Fe}]_{\rm 1P}\) shows no significant correlation with the main
orbital quantities, and the \(S_{\rm ChM,z}\)--\(z_{\max}\) anticorrelation
persists when it is added as a control. Although limited by the small overlap
sample, this check argues against the environmental trend being simply a hidden
baseline \(\alpha\)-abundance effect.

The second check concerns age. If clusters on different orbits also formed at
systematically different times, age could in principle mediate the observed
trend. We find no significant correlation between \(S_{\rm ChM,z}\) and cluster
age in the full sample or within the in-situ subsample. The GSE subsample shows
only a weak, statistically insignificant positive tendency, which is not interpretable given the small
number of clusters. Age is also uncorrelated with \(z_{\max}\), so it cannot explain the orbital trend.
Together, these tests indicate that the \(S_{\rm ChM,z}\)--orbit relation is
not driven by either baseline \(\alpha\)-chemistry or cluster age.

\end{document}